\documentclass[prl,twocolumn,aps,letterpaper, preprintnumbers,superscriptaddress]{revtex4-1}
\usepackage{amsthm,amsmath,amssymb}

\usepackage{mathrsfs}
\usepackage{amsthm}
\usepackage{amsmath}
\usepackage{amssymb}
\usepackage{graphicx}
\usepackage{epstopdf}
\usepackage{url}
\usepackage{subfigure}
\usepackage{float}
\usepackage{bm}
\usepackage{ifthen}
\usepackage[usenames,dvipsnames]{color}
\usepackage{mathrsfs}
\usepackage[colorlinks=true,citecolor=blue,urlcolor=black]{hyperref}
\usepackage{float}

\newcommand{\be}{\begin{equation}}
	\newcommand{\ee}{\end{equation}}

\newcommand{\sket}[1]{{\ensuremath{\lvert#1\rangle}}}
\newcommand{\lket}[1]{{\ensuremath{\left\lvert#1\right\rangle}}}
\newcommand{\ket}[1]{\if@display\lket{#1}\else\sket{#1}\fi}

\newcommand{\sbra}[1]{{\ensuremath{\langle#1\rvert}}}
\newcommand{\lbra}[1]{{\ensuremath{\left\langle#1\right\rvert}}}
\newcommand{\bra}[1]{\if@display\lbra{#1}\else\sbra{#1}\fi}

\newcommand{\sbraket}[2]{{\ensuremath{\langle#1\rvert#2\rangle}}}
\newcommand{\lbraket}[2]{{\ensuremath{\left\langle#1\!\left\rvert\vphantom{#1}#2\right.\!\right\rangle}}}
\newcommand{\braket}[2]{\if@display\lbraket{#1}{#2}\else\sbraket{#1}{#2}\fi}

\newcommand{\sketbra}[2]{{\ensuremath{\lvert #1\rangle\!\langle #2\rvert}}}
\newcommand{\lketbra}[2]{{\ensuremath{\left\lvert #1\right\rangle\!\!\left\langle #2\right\rvert}}}
\newcommand{\ketbra}[2]{\if@display\lketbra{#1}{#2}\else\sketbra{#1}{#2}\fi}

\theoremstyle{plain}

\theoremstyle{definition}

\begin{document}

\title{Unconditionally secure key distribution without quantum channel}

\author{Hua-Lei Yin}\email{hlyin@ruc.edu.cn}
\affiliation{Department of Physics and Beijing Key Laboratory of Opto-electronic Functional Materials and Micro-nano Devices, Key Laboratory of Quantum State Construction and Manipulation (Ministry of Education), Renmin University of China, Beijing 100872, China}
\date{\today}

\begin{abstract}
Key distribution plays a fundamental role in cryptography~\cite{menezes2018handbook,stinson1995cryptography,goldreich2006foundations,yin2023experimental}. Currently, the quantum scheme~\cite{bennett1984quantum,ekert1991quantum} stands as the only known method for achieving unconditionally secure key distribution. This method has been demonstrated over distances of 508 and 1002 kilometers in the measurement-device-independent~\cite{zhou2023experimental}  and twin-field~\cite{liu2023experimental} configurations, respectively. However, quantum key distribution faces transmission distance issues~\cite{diamanti2016practical} and numerous side channel attacks~\cite{xu2020secure} since the basic physical picture requires the use of quantum channels between users~\cite{xu2020secure,gisin2002quantum,scarani2009security,portmann2022security}. Even when quantum repeater~\cite{Sangouard2011quantum} and quantum constellation~\cite{lu2022micius} are used, commercializing quantum cryptography on a large scale remains unattainable due to the considerable expense and significant technical hurdles associated with establishing a global quantum network and facilitating mobile quantum communication. Here, by discovering the provable quantum one-way function, we propose another key distribution scheme with unconditional security, named probability key distribution, that promises users between any two distances to generate a fixed and high secret key rate. There are no quantum channels for exchanging quantum signals between two legitimate users. Non-local entangled states can be generated, identified and measured in the equivalent virtual protocol and can be used to extract secret keys. We anticipate that this discovery presents a paradigm shift in achieving unconditionally secure cryptography, thereby facilitating its widespread application on a global scale.
\end{abstract}

\maketitle


Cryptography ensures the secure confidentiality, integrity, authenticity, and non-repudiation during data processing~\cite{menezes2018handbook,stinson1995cryptography,goldreich2006foundations,yin2023experimental}.
Modern cryptography ensures security through keys rather than unknown cryptosystems. How to distribute secret keys in the presence of an eavesdropper Eve is known as the key distribution problem.
The holy grail of key distribution is unconditional security, also called information-theoretic security. To present our protocol, we outline several typical key distribution schemes in Table~\ref{Tab1}. Presently, public-key cryptography~\cite{menezes2018handbook} is the most extensively adopted and efficacious scheme, primarily due to its efficiency and no requirement on a physical channel. Unfortunately, public-key cryptography is very vulnerable to quantum computing~\cite{shor1999polynomial}. Recently, post-quantum cryptography has been proposed to replace public-key cryptography by introducing greater mathematical complexity~\cite{bernstein2017post}. However, this approach can only resist some known quantum attacks and cannot ensure security against future advanced algorithms~\cite{MATZOV2022Report}. An alternative approach to resolving the key distribution predicament involves using physical laws, such as chaos key distribution~\cite{argyris2005chaos,Soriano2013complex}, optical (quantum) stream cipher~\cite{Barbosa2003secure}, and quantum key distribution~\cite{bennett1984quantum,ekert1991quantum}.
These physics-based key distribution schemes require physics channels to exchange physics signals.  Consequently, the key rate and transmission distance are limited by channel loss, and side channel attacks via physics channels are inevitable.

\begin{table*}[t]
\centering
\caption{A comparison of several typical key distribution schemes.} \label{Tab1}
\begin{tabular}{c|c|c|c|c}
  \hline \hline
   Scheme & Fundamental laws &  Physics channel & Key rate & Unconditional security\\
  \hline
  Public-key cryptography & Computation complexity& No & Low & No\\
  \hline
  Post-quantum cryptography & Computation complexity& No & Low & No \\
  \hline
  Chaos key distribution & Chaos synchronization & Yes & High & No \\
  \hline
  Optical stream cipher & Semi-classical physics & Yes & High & No \\
  \hline
  Quantum key distribution & Quantum physics  & Yes & Low & Yes\\
  \hline
  Probability key distribution & Probability theory& No & High & Yes \\
         & and quantum physics & & & \\
\hline  \hline
\end{tabular}
\end{table*}

Quantum key distribution has garnered significant attention
and advancement
due to its exclusive capability to ensure unconditional security through rigorous theoretical analysis~\cite{gisin2002quantum,scarani2009security,xu2020secure,portmann2022security}.
For instance, a space-to-ground quantum network spanning 4600 kilometers has been successfully demonstrated leveraging a quantum satellite~\cite{chen2021integrated}.
Since the inception of the initial quantum protocol by Bennett and Brassard in 1984~\cite{bennett2014quantum}, the ingrained physical picture involves two distant users, Alice and Bob, engaging a quantum channel for the exchange of quantum signals~\cite{gisin2002quantum,scarani2009security,xu2020secure,portmann2022security}. Loosely speaking, in an actual protocol or at least an equivalent virtual protocol, quantum entanglement will establish between Alice and Bob through the exchange of quantum signals. Subsequently, this entanglement serves as the foundational resource for Alice and Bob to distill a secure key via quantum laws. Owing to the inevitable loss in the quantum channel, the secret key rate are limited by the capacity of the quantum channel, such as the repeaterless bound~\cite{takeoka2014fundamental,pirandola2017fundamental} and the repeater-assisted bound~\cite{pirandola2019end}. Envisioning a prospective global quantum communication network involves interlinking ground-based nodes via quantum repeater~\cite{Sangouard2011quantum}  and connecting satellite-ground and inter-satellite nodes through a quantum constellation~\cite{lu2022micius}.
However, quantum repeaters, for example, require a combination of efficient and high-fidelity quantum memories, gate operations, and measurements, remain an outstanding challenge.
Quantum constellations not only have security compromises but also require extremely high technical difficulty and substantial financial investment.

\begin{figure*}[t]
\centering
\includegraphics[width=18cm]{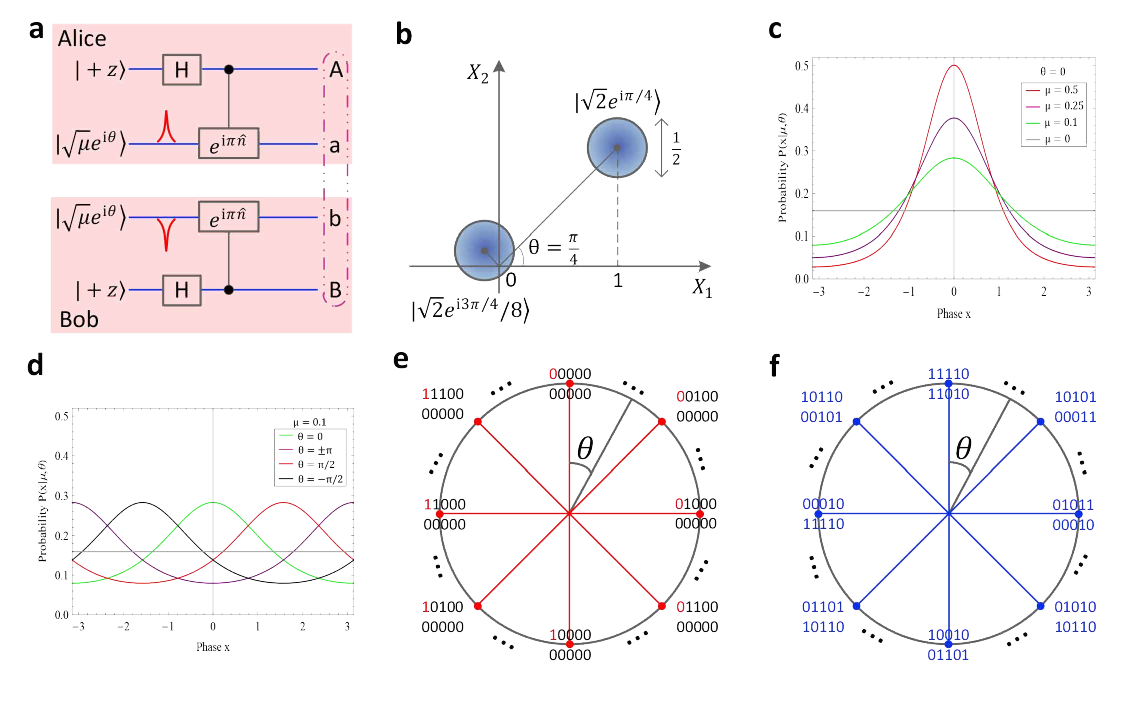}
\caption{\textbf{Basic idea behind non-local entangled state generation.} \textbf{a}, Alice (Bob) exploits a Hadamard gate to qubit $A$ ($B$) and a controlled-phase gate $\ket{+z}\bra{+z}\otimes \hat{\textbf{I}}+\ket{-z}\bra{-z}\otimes e^{\textbf{i}\pi \hat{n}}$ to qubit $A$ ($B$) as a control and optical pulse $a$ ($b$) as a target. If the global phase $\theta$ of Alice's and Bob's optical pulses are identical and randomized, qubit $A$ and qubit $B$ will be entangled from the view of Eve. \textbf{b}, Fluctuations of the quadrature operators $\hat{X}_{1}=(\hat{a}^{\dagger}+\hat{a})/2$ and $\hat{X}_{2}=\textbf{i}(\hat{a}^{\dagger}-\hat{a})/2$ of the coherent state in phase space. This means that given a coherent state, the measured phase will fluctuate.
\textbf{c}, \textbf{d}, Phase-probability distributions of coherent states with different intensities and global phases. It is clearly observed that the higher the intensity of the coherent state is, the smaller the variance of the phase. The gray line represents the vacuum state where the phase is uniformly random and the probability is equal to $\frac{1}{2\pi}$. \textbf{e}, Traditional mapping rule, where Eve knows the mapping rule and is fixed in all sessions. The global phase is directly generated according to the value of the 10-bit string $\vec{x}$, i.e., $\theta=2\pi j/1024$, where $j$ is the decimal value of the 10-bit string $\vec{x}$. The 10-bit $\vec{x}=1111011010$ ($j=986$) corresponds to the $\theta=2\pi\times986/1024$ phase in all the sessions. \textbf{f}, Random mapping rule in one session, where Eve has no knowledge of the mapping rule and is changed in each session. The 10-bit $\vec{x}=1111011010$ corresponds to phase $\theta=0$ in this special session.
} \label{f1}
\end{figure*}

\subsection{Non-local entangled state generation}

The key distribution scheme delineated herein not only pledges unconditional security but also obviates the necessity for a physical (quantum) channel. It promises Alice and Bob in establishing a consistent and high secret key rate irrespective of distance. Prior to the specific introduction of our probability key distribution (PKD) protocol, we initially elucidate the generation of non-local entangled states, depicted in Fig.~\ref{f1}a. The utilization of the Hadamard gate and controlled-phase gate generates an entangled state $\frac{1}{\sqrt{2}}\left(\ket{+z}\ket{\sqrt{\mu}e^{\textbf{i}\theta}}+\ket{-z}\ket{-\sqrt{\mu}e^{\textbf{i}\theta}}\right)$ between the qubit and optical pulse, wherein $\ket{\pm z}$ represents the eigenstates of the $Z$ basis. Considering the case of continuous phase randomization, the joint density matrix $\hat{\rho}$ of Alice's and Bob's systems is a mixture of states $\hat{\rho}_{k}$ (see Methods).
Qubit systems $A$ and $B$ are entangled, and there is no phase shift error for each photon number $k$. Evidently, the virtual entangled state is non-locally generated as both Alice and Bob possess identical and confidential random phase information. Leveraging the virtual entangled state enables Alice and Bob to extract a secret key. Now, we will elucidate how Alice and Bob can continuously exchange secret phase information $\theta$ with unconditional security by employing the following two pivotal observations.

\subsection{Provable quantum one-way function}

\emph{First observation}: The process of generating the discrete phase-randomized weak coherent state through a random mapping rule is the provable quantum one-way function if the phase number is sufficiently large and the optical intensity remains low. Let us define our provable \emph{quantum one-way function}, which can be regarded as the quantum version of a one-way function~\cite{goldreich2006foundations} and has rigorous one-wayness.
A quantum function $qf:\vec{x}\in\{0,1\}^{l}\rightarrow\ket{\phi_{j}(\vec{x})}$ is called provable quantum one-way if the following two conditions hold: 1) easy to evaluate, enabling the generation of a quantum state $\ket{\phi_{j}(\vec{x})}$ in polynomial time corresponding to the input bit string $\vec{x}$; 2) \emph{unable to invert}, preventing the derivation of any meaningful information about the bit string $\vec{x}$ from the received quantum state (Note that the received quantum state should be regarded as the mixture state instead of pure state $\ket{\phi_{j}(\vec{x})}$ due to missing information $\vec{x}$).

The continuous phase-randomized weak coherent state can be seen as a mixture of photon-number states~\cite{van2001Quantum}
\begin{equation}
\begin{aligned}\label{eq4}
\int_{0}^{2\pi}\frac{d\theta}{2\pi}\ket{\sqrt{\mu}e^{\textbf{i}\theta}}\bra{\sqrt{\mu}e^{\textbf{i}\theta}}=\sum_{k=0}^{\infty}e^{-\mu}\frac{\mu^{k}}{k!}\ket{k}\bra{k},
\end{aligned}
\end{equation}
if Eve is not known about the phase $\theta$ of each coherent state. From the view of the right-hand side of Eq.~\eqref{eq4}, the process of generating the continuous phase-randomized coherent state is a provable quantum one-way function, where the photon number state hides the global phase information $\theta$. The main reasons are that there are quantum fluctuations for quadrature operators of the coherent state in phase space, as shown in Fig.~\ref{f1}b. The phase-probability distributions of the coherent state $\ket{\sqrt{\mu}e^{\textbf{i}\theta}}$ are shown in Figs.~\ref{f1}c and \ref{f1}d with different intensities $\mu$ and phases $\theta$, respectively. Measuring the phase of a continuous phase-randomized coherent state equates to measuring the Poisson-distributed mixture of the photon number state (see Methods).

The continuous phase randomization scenarios entail an infinite number of global phases. However, it is essential to consider discrete phase randomization, where one has $m$ global phases and $\theta\in{0,2\pi/m,4\pi/m,\ldots,2\pi(m-1)/m}$. We must note the important difference between the continuous and discrete phase-randomized cases. As demonstrated in Fig~\ref{f1}e, mapping a global phase needs $|\vec{x}|=\log_{2}m=10$ bits if $m=1024$. In the traditional mapping rule, the first bits of $\vec{x}$ are always $0$ and $1$ for $\theta\in[0,\pi)$ and $\theta\in[\pi,2\pi)$, respectively. Noteworthy is the non-uniform phase-probability distribution of the coherent state ($\mu\neq0$), as depicted in Fig~\ref{f1}c. Eve can directly deduce, with high probability, the values of the first bit as 0 and 1 when the measured global phase corresponds to $\pi/2$ and $3\pi/2$, respectively. Consequently, the discrete phase-randomized coherent state even with $m=1024$ is not a rigorous quantum one-way function if the traditional mapping rule is utilized. Note that this will become the rigorous quantum one-way function if $m\rightarrow\infty$ because a few bits of information is no meaning for an infinitely long bit string $|\vec{x}|=\log_{2}m\rightarrow\infty$. Here, we eliminate the probability difference of the bit value by exploiting the \emph{random mapping} rule, as shown in Fig.~\ref{f1}f. The phase $\theta=0$ does not just correspond to only one 10-bit string $\vec{x}=0000000000$ anymore but to all 1024 feasible 10-bit strings $\vec{x}\in\{0,1\}^{10}$.
Each global phase uniformly and randomly correlates with all possible $10$-bit strings. Hence, under the random mapping rule, Eve can not steal any information of $\vec{x}$ even he has the discrete phase-randomized weak coherent state (details see Supplementary Information).

\subsection{Random information negotiation}

\emph{Second observation}: Alice and Bob can share themselves-generated quantum random numbers in many communication rounds with unconditional security via a fixed but long secret key and a per-update but short secret key if these quantum random numbers are not leaked to Eve when they are used. Let us define the unconditional security of a communication round where Alice and Bob transmit ciphertext to share plaintext via key and algorithm over an authentic channel. We call the communication round unconditionally secure when Eve cannot steal any plaintext information even if she uses unlimited computational resources. To elaborate, Eve can merely guess the plaintext at random; even if she guessed the plaintext correctly, she could not find any difference between the correct plaintext and the other wrong plaintext.

For an encryption system, it has been proven that one-time pad can provide unconditional security~\cite{shannon1949communication}, where the ciphertext $\vec{c}$ is the XOR value between the plaintext $\vec{m}$ and the key $\vec{k}$, i.e., $\vec{c}=\vec{m}\oplus \vec{k}$. For one-time pad, the key $\vec{k}$ should be completely random and up-dated in each round (i.e., one key can only used once) while there is no any requirement on the plaintext for different rounds~\cite{stinson1995cryptography}.
Here we exchange the above requirements of the key and plaintext in our random information negotiation. Let plaintext be a quantum random number $\vec{r}$, which is completely random and up-dated in different round. Let key $\vec{d}$ is generated by using a fixed but long secret key $\vec{K}_{\rm fix}$ and a up-date but short secret key $\vec{K}_{\rm upd}$ in each round. The length of the fixed secret key $\vec{K}_{\rm fix}$ can not be shorter than $|\vec{d}|$, which ensures that the generated key $\vec{d}$ is completely random in the first round. Then Alice and Bob utilize the generated key $\vec{d}$ to share the quantum random number $\vec{r}$ via XOR algorithm $\vec{c}=\vec{r}\oplus\vec{d}$. If we can ensure that the used quantum random number $\vec{r}$ does not leak any information to Eve (it is true in provable quantum one-way function), who can only implement ciphertext-only attack and no other plaintext-dependent attacks. Therefore, the shared quantum random number $\vec{r}$ and key $\vec{d}$ (as well as key $\vec{K}_{\rm fix}$) are all unknown to Eve in the first communication round since observation of the ciphertext provides no plaintext information whatsoever to Eve. Based on this, Alice and Bob can repeat the above process in many communication rounds, where the unconditional security is always maintain even key $\vec{K}_{\rm fix}$ is reused (see Supplementary Information for detail).

\begin{figure*}[t]
\centering
\includegraphics[width=18cm]{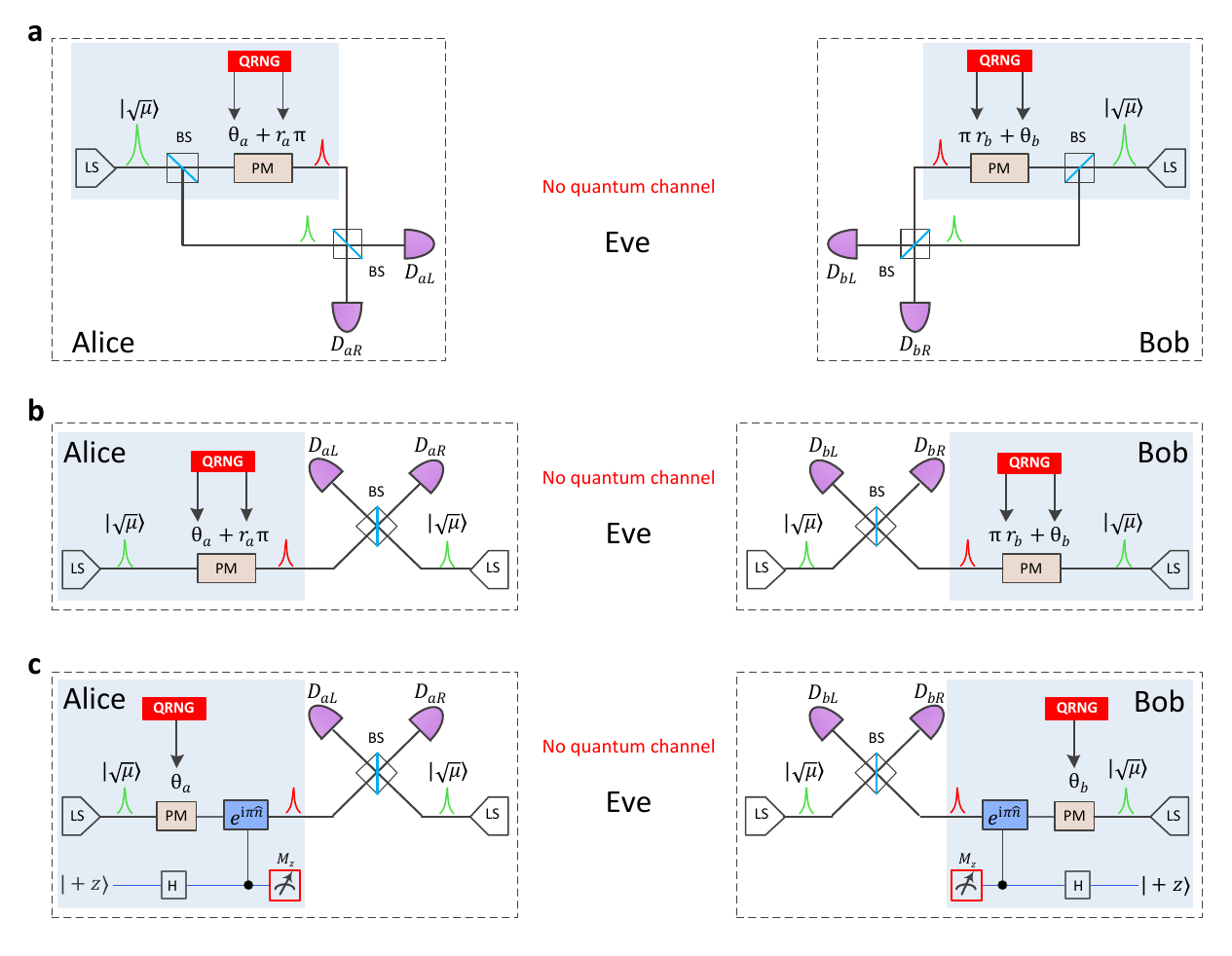}
\caption{\textbf{Optical realization of the PKD protocol. a}, Actual protocol: Alice (Bob) exploits a laser (LS) and a beam splitter (BS) to generate optical pulse pairs (called the signal pulse and reference pulse, respectively). She (he) utilizes a phase modulator (PM) to modulate the phase of the signal pulse $\theta_{a}+r_{a}\pi$ ($\theta_{b}+r_{b}\pi$), where $\theta_{a}$ ($\theta_{b}$) are random global phases and $r_{a},r_{b}$ are binary random numbers generated by quantum random number generation (QRNG). Then, Alice (Bob) performs single-photon interference measurements for the pulse pairs through a BS and two detectors, $D_{aL}$ and $D_{aR}$ ($D_{bL}$ and $D_{bR}$, respectively). \textbf{b}, Virtual protocol 1: Alice (Bob) exploits one laser to prepare the signal pulse and another laser to generate the reference pulse. \textbf{c}, Virtual protocol 2: Alice (Bob) generates an entangled state between the qubit and optical modes instead of an signal pulse. Alice and Bob measure their qubit to obtain the raw key by using the $Z$ basis after they announce the successful detection event.
} \label{f2}
\end{figure*}

\subsection{Probability key distribution protocol}

Based on the above two observations, we propose our actual PKD protocol, as shown in Fig.~\ref{f2}a. The fixed but long secret key $\vec{K}_{\rm fix}$ is reused many times before it is discarded. Alice and Bob exploit $\vec{K}_{\rm fix}\in\{0,1\}^{s+t-1}$ to build the Toeplitz matrix $\textbf{H}_{st}$ with $s$ rows and $t$ columns. The per-update but short secret key $\vec{K}\in\{0,1\}^{s}$ will change in each PKD session. For each PKD session, five steps were performed as follows.

(i) Alice prepares weak coherent state optical pulse pairs (signal pulse and reference pulse) $\ket{\sqrt{\mu}e^{\textbf{i}\varphi_{a}}}_{a}\otimes\ket{\sqrt{\mu}}_{a}$, where $\mu$ is the intensity of each pulse and $\varphi_{a}=\theta_{a}+r_{a}\pi$ is the phase of the signal pulse. $r_{a}\in\{0,1\}$ is the random bit value. The random global phase $\theta_{a}\in\{0,2\pi/m,4\pi/m,\ldots,2\pi(m-1)/m\}$ is determined by the quantum random number string $\vec{x}_{a}\in\{0,1\}^{\log_{2}m}$. There are $m$ global phases and each $\log_{2}m$-bit map to a global phase. Bob does the same. Instead of the fixed mapping rule, the global phase is determined by random mapping. The random mapping rule is shared between Alice and Bob through one-time pad by consuming $m\log_{2}m$ bits of the pre-shared secret key. They repeat step (i) for $N$ rounds.
(ii) Alice performs single-photon interference measurement for the prepared pulse pairs. If and only if one of detectors $D_{aL}$ and $D_{aR}$ clicks represents a successful detection event. Alice keeps $r_{a}$ as the raw key and announces the successful detection event and the corresponding detector when there is a successful detection event. Bob does the same. The numbers of successful detection events for Alice and Bob are both $n$ for one PKD session.
(iii) Alice and Bob obtain the data string $\vec{D}=\vec{K}_{\rm upd}\cdot\textbf{H}_{st}$, where $\vec{D}$ has $t$ bits and $t\geq n\log_{2}m$. Let the first $n\log_{2}m$ bits in $\vec{D}$ constitute the data string $\vec{D}_{n}$. For $n$ successful detection events of Alice, let $\vec{R}_{n}(\vec{x}_{a})\in\{0,1\}^{n\log_{2}m}$ be the random bit string corresponding to Alice's global phases. Alice sends ciphertext $\vec{R}_{n}(\vec{x}_{a})\oplus \vec{D}_{n}$ to Bob, who decrypts ciphertext with data string $\vec{D}_{n}$ to obtain quantum random numbers $\vec{R}_{n}(\vec{x}_{a})$ and thus acquires the global phase of Alice $\theta_{a}$ according to the above random mapping rule. According to the phase $\theta_{a}$ of each event, Bob rearranges his raw key and detector click order to ensure that $\theta_{b}=\theta_{a}$. If $\{D_{aL},D_{bR}\}$ or $\{D_{aR},D_{bL}\}$ click, Bob flips his raw key bit. Finally, Alice and Bob obtain the raw key strings $\vec{Z}_{a}$ and $\vec{Z}_{b}$, respectively.
(iv) Bob acquires an estimation $\vec{\textrm{Z}}_{b}$ of $\vec{Z}_{a}$ in the error correction scheme by revealing at most $\lambda$ bits of information. Then, Alice and Bob perform an error verification to ensure identical keys by publishing $\log_{2}(2/\varepsilon_{\rm cor})$-bit~\cite{wegman1981new, krawczyk1994lfsr}.
(v) Alice and Bob perform privacy amplification by applying a random universal$_2$ hash function~\cite{bennett1995generalized} to extract length $\ell$ bits of secret key.

Note that the sharing $\vec{R}_{n}(\vec{x}_{a})$ scheme in step (iii) is a special way in our random information negotiation.
For the actual protocol, the global phase of the reference pulse can be known to Eve.
Thus, Alice (Bob) can utilize another laser to prepare the reference pulse with zero phase and performs the single-photon interference measurement, which can be called the prepare-and-measure virtual protocol 1, as shown in Fig.~\ref{f2}b.
Furthermore, the entanglement-based virtual protocol 2 in Fig.~\ref{f2}c is equivalent to the prepare-and-measure virtual protocol 1. The phase error of the joint qubit system of Alice and Bob is always zero (see Methods).

The PKD protocol is $\varepsilon_{\rm sec}$-secret and $\varepsilon_{\rm cor}$-correct if the secret key length of one session is satisfied (see the Supplementary Information).
\begin{equation}
\begin{aligned}\label{eq2}
\ell \leq n-\lambda-\log_{2}\frac{2}{\varepsilon_{\rm cor}}-2\log_{2}\frac{3}{2\varepsilon_{\rm sec}},
\end{aligned}
\end{equation}
where the leaked information is $\lambda=nfh(E)$ due to error correction. The Shannon entropy function is $h(x)=-x\log_{2}x-(1-x)\log_{2}(1-x)$. $E\approx25\%$ and $f=1.05$ are the bit error rate and the error correction efficiency, respectively. Here, we can set the intensity $\mu=0.1$, the phase number $m=1024$, the detection efficiency $\eta_{d}=0.8$, the dark count rate $p_{d}=10^{-8}$, $\varepsilon_{\rm cor}=10^{-15}$ and $\varepsilon_{\rm sec}=10^{-10}$. Considering a gigahertz system and $N=10^{9}$ for one second in one PKD session, the net remaining secret key rate is $R=\ell-s-m\log_{2}m\approx20$ Mbps if we ignore the reused secret key string $\vec{K}_{\rm fix}$ and let $|\vec{K}_{\rm upd}|=s=10^{4}$.

\subsection{Conclusion and discussion}

Overall, we have proven the unconditional security of PKD by introducing the concepts of a rigorous quantum one-way function and random information negotiation. Our proposal allows any two distant users to extract secure keys at a high-speed rate as long as they can pre-share some of the secret keys. Based on the random mapping rule, our provable quantum one-way function does not reveal even a single bit information to Eve, which is completely different from the previous definition of the quantum one-way function~\cite{buhrman2001quantum}. We remark that our random information negotiation is not an encryption scheme since the communication data between Alice and Bob are quantum random numbers (generated by only Alice and Bob themselves) rather than messages. Moreover, shared quantum random numbers cannot be used as keys for encrypting another message. Otherwise, quantum random numbers could potentially be leaked to Eve, which leads to our random information negotiation that is not secure. Fortunately, shared quantum random numbers can be used to map the global phase of the coherent state when one constructs the quantum one-way function. These quantum random numbers do not leak to Eve due to the rigorous one-wayness inherent in our provable quantum one-way function.

The existence of classical one-way functions remains an unresolved inquiry. Should they exist, this can resolve the paramount unsolved query within theoretical computer science, namely, the complexity classes $P$ and $NP$ are distinct~\cite{goldreich2006foundations}. We have proven that the phase-randomized weak coherent state with a random mapping rule is a provable quantum one-way function with rigorous one-wayness. This function safeguards the phase information within quantum states, rendering it unrecoverable by Eve. As an example, we utilized this provable quantum one-way function to propose an unconditionally secure key distribution scheme. We expect that our quantum one-way function will be widely used to construct information-theoretically secure privacy protection protocols, such as quantum zero-knowledge proof and quantum secure multiparty computation. Actually, the steps required to prepare and measure quantum state within our PKD protocol are unnecessary and it will greatly simplify the cost and increase the bit rate, which will be discussed in our next work.

\section{Methods}

\subsection{Joint density matrix}

For continuous phase randomization, the joint density matrix (including qubit systems $A$, $B$ and optical modes $a$, $b$) is $\hat{\rho}=\sum_{k=0}^{\infty}\frac{e^{-2\mu}(2\mu)^{k}}{k! }\hat{\rho}_{k}$, where the state $\hat{\rho}_{k}$ can be given by
\begin{equation}\label{eq3}
   \begin{cases}
   \hat{P}\left(\frac{\ket{\phi^{-}}_{AB}\ket{+}_{ab}^{\otimes k}+\ket{\psi^{-}}_{AB}\ket{-}_{ab}^{\otimes k}}{\sqrt{2}}\right),&\mbox{if $k$ is odd},\\
   \\
  \hat{P}\left(\frac{\ket{\phi^{+}}_{AB}\ket{+}_{ab}^{\otimes k}+\ket{\psi^{+}}_{AB}\ket{-}_{ab}^{\otimes k}}{\sqrt{2}}\right),&\mbox{if $k$ is even}.
   \end{cases}
  \end{equation}
Let $\hat{P}(\ket{x})=\ket{x}\bra{x}$, states $\ket{\phi^{\pm}}=\frac{1}{\sqrt{2}}(\ket{+z}\ket{+z}\pm\ket{-z}\ket{-z})$ and  $\ket{\psi^{\pm}}=\frac{1}{\sqrt{2}}(\ket{+z}\ket{-z}\pm\ket{-z}\ket{+z})$ be four Bell states. States $\ket{\pm}_{ab}=\frac{1}{\sqrt{2}}(\ket{10}_{ab}\pm\ket{01}_{ab})=\frac{\hat{a}^{\dagger}\pm \hat{b}^{\dagger}}{\sqrt{2}}\ket{00}_{ab}$ is a superposition single-photon state with $a$ and $b$ modes. $\ket{\pm}_{ab}^{\otimes k}=\frac{1}{\sqrt{2^{k}k! }}\left(\hat{a}^{\dag}\pm \hat{b}^{\dag}\right)^{k}\ket{00}_{ab}$ is the $k$-photon state, i.e., $k$ identical photons with state $\ket{\pm}_{ab}$.

For the case of discrete phase randomization, the initial joint density matrix of Alice's and Bob's can be written as
\begin{equation}
\begin{aligned}
\hat{\rho}^{\rm dis}(m)=\sum_{k=0}^{m-1}P^{2\mu}_{m}(k)\hat{\rho}_{\lambda_{k}},
\end{aligned}
\end{equation}
which is the mixture of states $\hat{\rho}_{\lambda_{k}}$ with probability $P^{2\mu}_{m}(k)=e^{-2\mu}\sum_{l=0}^{\infty}\frac{(2\mu)^{lm+k}}{(lm+k)! }$. The density matrix of $\hat{\rho}_{\lambda_{k}}$ can be given by
\begin{equation}\label{eq5}
\begin{cases}
\hat{P}\left(\frac{\ket{\phi^{-}}_{AB}\ket{\lambda_{k}^{+}}_{ab}
+\ket{\psi^{-}}_{AB}\ket{\lambda_{k}^{-}}_{ab}}{\sqrt{2}} \right),  &\mbox{if $k$ is odd},\\
\\
\hat{P}\left(\frac{\ket{\phi^{+}}_{AB}\ket{\lambda_{k}^{+}}_{ab}+\ket{\psi^{+}}_{AB}\ket{\lambda_{k}^{-}}_{ab}}{\sqrt{2}}\right), &\mbox{if $k$ is even},
\end{cases}
\end{equation}
where the quantum state $\ket{\lambda_{k}^{\pm}}_{ab}$ is denoted as
\begin{equation}
\begin{aligned}
\ket{\lambda_{k}^{\pm}}_{ab}&=\frac{e^{-\mu}}{\sqrt{P^{2\mu}_{m}(k)}}\sum_{l=0}^{\infty}\frac{(\sqrt{2\mu})^{lm+k}}{\sqrt{(lm+k)! }}\ket{\pm}_{ab}^{\otimes lm+k}.\\
\end{aligned}
\end{equation}
If the phase number $m$ approaches infinity, the discrete phase randomization case will become a continuous case. According to Eqs.~\eqref{eq3} and~\eqref{eq5}, two qubit systems $A$ and $B$ can only be in the subspace spanned by Bell state $\ket{\phi^{-}}$  and $\ket{\psi^{-}}$ ($\ket{\phi^{+}}$ and $\ket{\phi^{+}}$) if $k$ is odd (even), no matter whether Eve performs any quantum measurements for optical modes $a$ and $b$. Obviously, there may have bit error but not phase error when Alice and Bob measure the qubit systems $A$ and $B$, respectively.

\subsection{Quantum measurement}
Given a coherent state with intensity $\mu$ and phase $\theta$, there is a phase-probability distribution that can be written as~\cite{Pegg:1989:Phase,Buzek:1992:Coherent}
\begin{equation}
\begin{aligned}
P(x|\mu,\theta)=\lim_{l\rightarrow\infty}\frac{1}{2\pi}\left|\sum_{k=0}^{l}e^{-\textbf{i}(x-\theta)k}\frac{e^{-\mu/2}\mu^{k/2}}{\sqrt{k! }}\right|^{2},
\end{aligned}
\end{equation}
where $x\in[0,2\pi)$ represents the measured phase, $P(x|\mu,\theta)$ is the corresponding probability and $\int_{0}^{2\pi}P(x|\mu,\theta)dx=1$.
The average probability of the measured phase $x$ for a given uniformly distributed phase $\theta\in[0,2\pi)$ is
\begin{equation}
\begin{aligned}
\frac{1}{2\pi}\int_{0}^{2\pi}P(x|\mu,\theta)d\theta\equiv\frac{1}{2\pi},~~~\forall~\mu~\textrm{and}~x,
\end{aligned}
\end{equation}
which means that the measured phase $x$ is completely random. If the intensity is zero (vacuum state), the phase-probability distribution is uniform, i.e., $P(x|\mu=0,\theta)=1/(2\pi)$. Actually, for all photon number states $\ket{k}$, the phase-probability distribution is uniform. Measuring the phase of a continuous phase-randomized coherent state is equivalent to measuring the mixture of the photon number state with the Poisson distribution.

For the discrete phase randomization $m=1024$ case, the average probability of the measured phase $x$ can be given by
\begin{equation}
\begin{aligned}
\frac{1}{1024}\sum_{j=0}^{1023}P\left(x|\mu=0.1,\theta=\frac{2\pi j}{1024}\right)\simeq\frac{1}{2\pi},~~~\forall~x,
\end{aligned}
\end{equation}
which is equivalent to the continuous phase randomization case with negligible difference in probability. The discrete phase-randomized coherent state can be written as a mixture of the pseudo photon-number state $\ket{\lambda_{k}}$~\cite{cao2015discrete, shao2023phase}
\begin{equation}
\begin{aligned}
\frac{1}{m}\sum_{j=0}^{m-1}\ket{\sqrt{\mu}e^{\textbf{i}\frac{2\pi j}{m}}}\bra{\sqrt{\mu}e^{\textbf{i}\frac{2\pi j}{m}}} = \sum_{k=0}^{m-1} P^{\mu}_{m}(k)\ket{\lambda_{k}}\bra{\lambda_{k}},
\end{aligned}
\end{equation}
where one has probability $P^{\mu}_{m}(k)=e^{-\mu}\sum_{l=0}^{\infty}\frac{\mu^{lm+k}}{(lm+k)! }$ and state $\ket{\lambda_{k}}=\frac{e^{-\mu/2}}{\sqrt{P^{\mu}_{m}(k)}}\sum_{l=0}^{\infty}\frac{(\sqrt{\mu})^{lm+k}}{\sqrt{(lm+k)!}}\ket{lm+k}$. The discrete case will become the continuous case if $m\rightarrow\infty$, i.e., $\lim_{m\rightarrow\infty}\ket{\lambda_{k}}\equiv\ket{k}$ and $\lim_{m\rightarrow\infty}P^{\mu}_{m}(k)\equiv e^{-\mu}\mu^{k}/k!$. For a microcosmic coherent state and a sufficiently large $m$, the discrete case is almost identical to the continuous case. The trace distance between $\ket{\lambda_{k}}$ and $\ket{k}$ can be given by
\begin{equation}
\begin{aligned}
D(\ket{\lambda_{k}},\ket{k})&=\frac{1}{2}{\rm tr}\big|\ket{\lambda_{k}}\bra{\lambda_{k}}-\ket{k}\bra{k}\big|\\
&<\sqrt{\sum_{l=1}^{\infty}\frac{\mu^{lm}k!}{(lm+k)! }}\approx\sqrt{\mu^{m}/m! }.
\end{aligned}
\end{equation}
For a discrete phase-randomized coherent state, the optimum unambiguous state discrimination measurement probability can be written as~\cite{chefles1998optimum,van2002unambiguous}
\begin{equation}
\begin{aligned}
P_{\rm USD}&=\min_{r=0,1,\ldots,m-1}\sum_{j=0}^{m-1}e^{-\textbf{i}2\pi jr/m}e^{\mu\left(e^{\textbf{i}2\pi j/m}-1\right)}\\
&\approx\frac{m\mu^{m-1}}{(m-1)! },
\end{aligned}
\end{equation}
where the second equation works if the intensity $\mu$ is small. If Eve implements the minimum error discrimination measurement, the minimum error probability is~\cite{barnett2009quantum,wallden2014minimum}
\begin{equation}
\begin{aligned}
P_{\rm min}=1-\frac{1}{m^{2}}\left|\sum_{r=0}^{m-1}\sqrt{\sum_{k=0}^{m-1}e^{-\mu(1-e^{\textbf{i}2\pi k/m})+\textbf{i}2\pi kr/m}}\right|^{2}.
\end{aligned}
\end{equation}
For the case of intensity $\mu=0.1$ and phase number $m=1024$, the trace distance $D(\ket{\lambda_{k}},\ket{k})\approx\sqrt{\mu^{m}/m!}=1.36\times10^{-1832}$. The optimum unambiguous state discrimination probability is $P_{\rm USD}\approx m\mu^{m-1}/(m-1)!=1.94\times10^{-3657}$, which means that Eve cannot successfully implement an unambiguous state discrimination attack. The minimum error probability is $P_{\rm min}=99.83\%$, which is approximately equal to the random guessing case in which the error probability is $1-1/m=99.90\%$.

\subsection{Acknowledgements}

We gratefully acknowledge the support from the National Natural Science Foundation of China (No. 12274223).


\begin{thebibliography}{0}%
\makeatletter
\providecommand \@ifxundefined [1]{%
 \@ifx{#1\undefined}
}%
\providecommand \@ifnum [1]{%
 \ifnum #1\expandafter \@firstoftwo
 \else \expandafter \@secondoftwo
 \fi
}%
\providecommand \@ifx [1]{%
 \ifx #1\expandafter \@firstoftwo
 \else \expandafter \@secondoftwo
 \fi
}%
\providecommand \natexlab [1]{#1}%
\providecommand \enquote  [1]{``#1''}%
\providecommand \bibnamefont  [1]{#1}%
\providecommand \bibfnamefont [1]{#1}%
\providecommand \citenamefont [1]{#1}%
\providecommand \href@noop [0]{\@secondoftwo}%
\providecommand \href [0]{\begingroup \@sanitize@url \@href}%
\providecommand \@href[1]{\@@startlink{#1}\@@href}%
\providecommand \@@href[1]{\endgroup#1\@@endlink}%
\providecommand \@sanitize@url [0]{\catcode `\\12\catcode `\$12\catcode `\&12\catcode `\#12\catcode `\^12\catcode `\_12\catcode `\%12\relax}%
\providecommand \@@startlink[1]{}%
\providecommand \@@endlink[0]{}%
\providecommand \url  [0]{\begingroup\@sanitize@url \@url }%
\providecommand \@url [1]{\endgroup\@href {#1}{\urlprefix }}%
\providecommand \urlprefix  [0]{URL }%
\providecommand \Eprint [0]{\href }%
\providecommand \doibase [0]{http://dx.doi.org/}%
\providecommand \selectlanguage [0]{\@gobble}%
\providecommand \bibinfo  [0]{\@secondoftwo}%
\providecommand \bibfield  [0]{\@secondoftwo}%
\providecommand \translation [1]{[#1]}%
\providecommand \BibitemOpen [0]{}%
\providecommand \bibitemStop [0]{}%
\providecommand \bibitemNoStop [0]{.\EOS\space}%
\providecommand \EOS [0]{\spacefactor3000\relax}%
\providecommand \BibitemShut  [1]{\csname bibitem#1\endcsname}%
\let\auto@bib@innerbib\@empty
\end{thebibliography}%


\begin{thebibliography}{10}
\expandafter\ifx\csname url\endcsname\relax
  \def\url#1{\texttt{#1}}\fi
\expandafter\ifx\csname urlprefix\endcsname\relax\def\urlprefix{URL }\fi
\providecommand{\bibinfo}[2]{#2}
\providecommand{\eprint}[2][]{\url{#2}}

\bibitem{menezes2018handbook}
\bibinfo{author}{Menezes, A.~J.}, \bibinfo{author}{van Oorschot, P.~C.} \& \bibinfo{author}{Vanstone, S.~A.}
\newblock \emph{\bibinfo{title}{Handbook of applied cryptography}} (\bibinfo{publisher}{CRC press}, \bibinfo{year}{1996}).

\bibitem{stinson1995cryptography}
\bibinfo{author}{Stinson, D.~R.}
\newblock \emph{\bibinfo{title}{Cryptography: theory and practice}} (\bibinfo{publisher}{CRC Press}, \bibinfo{year}{1995}).

\bibitem{goldreich2006foundations}
\bibinfo{author}{Goldreich, O.}
\newblock \emph{\bibinfo{title}{Foundations of Cryptography: Basic Techniques}} (\bibinfo{publisher}{Cambridge University Press}, \bibinfo{year}{2001}).

\bibitem{yin2023experimental}
\bibinfo{author}{Yin, H.-L.} \emph{et~al.}
\newblock \bibinfo{title}{Experimental quantum secure network with digital signatures and encryption}.
\newblock \emph{\bibinfo{journal}{Natl. Sci. Rev.}} \textbf{\bibinfo{volume}{10}}, \bibinfo{pages}{nwac228} (\bibinfo{year}{2023}).

\bibitem{bennett1984quantum}
\bibinfo{author}{Bennett, C.~H.} \& \bibinfo{author}{Brassard, G.}
\newblock \bibinfo{title}{Quantum cryptography: public key distribution and coin tossing}.
\newblock \emph{\bibinfo{journal}{In Proc. Int. Conf. on Computers, Systems and Signal Processing}} \bibinfo{pages}{175--179} (\bibinfo{year}{1984}).

\bibitem{ekert1991quantum}
\bibinfo{author}{Ekert, A.~K.}
\newblock \bibinfo{title}{Quantum cryptography based on bell's theorem}.
\newblock \emph{\bibinfo{journal}{Phys. Rev. Lett.}} \textbf{\bibinfo{volume}{67}}, \bibinfo{pages}{661--663} (\bibinfo{year}{1991}).

\bibitem{zhou2023experimental}
\bibinfo{author}{Zhou, L.} \emph{et~al.}
\newblock \bibinfo{title}{Experimental quantum communication overcomes the rate-loss limit without global phase tracking}.
\newblock \emph{\bibinfo{journal}{Phys. Rev. Lett.}} \textbf{\bibinfo{volume}{130}}, \bibinfo{pages}{250801} (\bibinfo{year}{2023}).

\bibitem{liu2023experimental}
\bibinfo{author}{Liu, Y.} \emph{et~al.}
\newblock \bibinfo{title}{Experimental twin-field quantum key distribution over 1000 km fiber distance}.
\newblock \emph{\bibinfo{journal}{Phys. Rev. Lett.}} \textbf{\bibinfo{volume}{130}}, \bibinfo{pages}{210801} (\bibinfo{year}{2023}).

\bibitem{diamanti2016practical}
\bibinfo{author}{Diamanti, E.}, \bibinfo{author}{Lo, H.-K.}, \bibinfo{author}{Qi, B.} \& \bibinfo{author}{Yuan, Z.}
\newblock \bibinfo{title}{Practical challenges in quantum key distribution}.
\newblock \emph{\bibinfo{journal}{npj Quantum Inf.}} \textbf{\bibinfo{volume}{2}}, \bibinfo{pages}{1--12} (\bibinfo{year}{2016}).

\bibitem{xu2020secure}
\bibinfo{author}{Xu, F.}, \bibinfo{author}{Ma, X.}, \bibinfo{author}{Zhang, Q.}, \bibinfo{author}{Lo, H.-K.} \& \bibinfo{author}{Pan, J.-W.}
\newblock \bibinfo{title}{Secure quantum key distribution with realistic devices}.
\newblock \emph{\bibinfo{journal}{Rev. Mod. Phys.}} \textbf{\bibinfo{volume}{92}}, \bibinfo{pages}{025002} (\bibinfo{year}{2020}).

\bibitem{gisin2002quantum}
\bibinfo{author}{Gisin, N.}, \bibinfo{author}{Ribordy, G.}, \bibinfo{author}{Tittel, W.} \& \bibinfo{author}{Zbinden, H.}
\newblock \bibinfo{title}{Quantum cryptography}.
\newblock \emph{\bibinfo{journal}{Rev. Mod. Phys.}} \textbf{\bibinfo{volume}{74}}, \bibinfo{pages}{145--195} (\bibinfo{year}{2002}).

\bibitem{scarani2009security}
\bibinfo{author}{Scarani, V.} \emph{et~al.}
\newblock \bibinfo{title}{The security of practical quantum key distribution}.
\newblock \emph{\bibinfo{journal}{Rev. Mod. Phys.}} \textbf{\bibinfo{volume}{81}}, \bibinfo{pages}{1301} (\bibinfo{year}{2009}).

\bibitem{portmann2022security}
\bibinfo{author}{Portmann, C.} \& \bibinfo{author}{Renner, R.}
\newblock \bibinfo{title}{Security in quantum cryptography}.
\newblock \emph{\bibinfo{journal}{Rev. Mod. Phys.}} \textbf{\bibinfo{volume}{94}}, \bibinfo{pages}{025008} (\bibinfo{year}{2022}).

\bibitem{Sangouard2011quantum}
\bibinfo{author}{Sangouard, N.}, \bibinfo{author}{Simon, C.}, \bibinfo{author}{de~Riedmatten, H.} \& \bibinfo{author}{Gisin, N.}
\newblock \bibinfo{title}{Quantum repeaters based on atomic ensembles and linear optics}.
\newblock \emph{\bibinfo{journal}{Rev. Mod. Phys.}} \textbf{\bibinfo{volume}{83}}, \bibinfo{pages}{33--80} (\bibinfo{year}{2011}).

\bibitem{lu2022micius}
\bibinfo{author}{Lu, C.-Y.}, \bibinfo{author}{Cao, Y.}, \bibinfo{author}{Peng, C.-Z.} \& \bibinfo{author}{Pan, J.-W.}
\newblock \bibinfo{title}{Micius quantum experiments in space}.
\newblock \emph{\bibinfo{journal}{Rev. Mod. Phys.}} \textbf{\bibinfo{volume}{94}}, \bibinfo{pages}{035001} (\bibinfo{year}{2022}).

\bibitem{shor1999polynomial}
\bibinfo{author}{Shor, P.~W.}
\newblock \bibinfo{title}{Polynomial-time algorithms for prime factorization and discrete logarithms on a quantum computer}.
\newblock \emph{\bibinfo{journal}{SIAM review}} \textbf{\bibinfo{volume}{41}}, \bibinfo{pages}{303--332} (\bibinfo{year}{1999}).

\bibitem{bernstein2017post}
\bibinfo{author}{Bernstein, D.~J.} \& \bibinfo{author}{Lange, T.}
\newblock \bibinfo{title}{Post-quantum cryptography}.
\newblock \emph{\bibinfo{journal}{Nature}} \textbf{\bibinfo{volume}{549}}, \bibinfo{pages}{188--194} (\bibinfo{year}{2017}).

\bibitem{MATZOV2022Report}
\bibinfo{author}{MATZOV}.
\newblock \bibinfo{title}{Report on the security of lwe: Improved dual lattice attack}.
\newblock \emph{\bibinfo{journal}{https://doi.org/10.5281/zenodo.6493704}}  (\bibinfo{year}{2022}).

\bibitem{argyris2005chaos}
\bibinfo{author}{Argyris, A.} \emph{et~al.}
\newblock \bibinfo{title}{Chaos-based communications at high bit rates using commercial fibre-optic links}.
\newblock \emph{\bibinfo{journal}{Nature}} \textbf{\bibinfo{volume}{438}}, \bibinfo{pages}{343--346} (\bibinfo{year}{2005}).

\bibitem{Soriano2013complex}
\bibinfo{author}{Soriano, M.~C.}, \bibinfo{author}{Garc\'{\i}a-Ojalvo, J.}, \bibinfo{author}{Mirasso, C.~R.} \& \bibinfo{author}{Fischer, I.}
\newblock \bibinfo{title}{Complex photonics: Dynamics and applications of delay-coupled semiconductors lasers}.
\newblock \emph{\bibinfo{journal}{Rev. Mod. Phys.}} \textbf{\bibinfo{volume}{85}}, \bibinfo{pages}{421--470} (\bibinfo{year}{2013}).

\bibitem{Barbosa2003secure}
\bibinfo{author}{Barbosa, G.~A.}, \bibinfo{author}{Corndorf, E.}, \bibinfo{author}{Kumar, P.} \& \bibinfo{author}{Yuen, H.~P.}
\newblock \bibinfo{title}{Secure communication using mesoscopic coherent states}.
\newblock \emph{\bibinfo{journal}{Phys. Rev. Lett.}} \textbf{\bibinfo{volume}{90}}, \bibinfo{pages}{227901} (\bibinfo{year}{2003}).

\bibitem{chen2021integrated}
\bibinfo{author}{Chen, Y.-A.} \emph{et~al.}
\newblock \bibinfo{title}{An integrated space-to-ground quantum communication network over 4,600 kilometres}.
\newblock \emph{\bibinfo{journal}{Nature}} \textbf{\bibinfo{volume}{589}}, \bibinfo{pages}{214--219} (\bibinfo{year}{2021}).

\bibitem{bennett2014quantum}
\bibinfo{author}{Bennett, C.~H.} \& \bibinfo{author}{Brassard, G.}
\newblock \bibinfo{title}{Quantum cryptography: Public key distribution and coin tossing?}
\newblock \emph{\bibinfo{journal}{Theor. Comput. Sci.}} \textbf{\bibinfo{volume}{560}}, \bibinfo{pages}{7--11} (\bibinfo{year}{2014}).

\bibitem{takeoka2014fundamental}
\bibinfo{author}{Takeoka, M.}, \bibinfo{author}{Guha, S.} \& \bibinfo{author}{Wilde, M.~M.}
\newblock \bibinfo{title}{Fundamental rate-loss tradeoff for optical quantum key distribution}.
\newblock \emph{\bibinfo{journal}{Nat. Commun.}} \textbf{\bibinfo{volume}{5}}, \bibinfo{pages}{5235} (\bibinfo{year}{2014}).

\bibitem{pirandola2017fundamental}
\bibinfo{author}{Pirandola, S.}, \bibinfo{author}{Laurenza, R.}, \bibinfo{author}{Ottaviani, C.} \& \bibinfo{author}{Banchi, L.}
\newblock \bibinfo{title}{Fundamental limits of repeaterless quantum communications}.
\newblock \emph{\bibinfo{journal}{Nature Commun.}} \textbf{\bibinfo{volume}{8}}, \bibinfo{pages}{15043} (\bibinfo{year}{2017}).

\bibitem{pirandola2019end}
\bibinfo{author}{Pirandola, S.}
\newblock \bibinfo{title}{End-to-end capacities of a quantum communication network}.
\newblock \emph{\bibinfo{journal}{Commun. Phys.}} \textbf{\bibinfo{volume}{2}}, \bibinfo{pages}{51} (\bibinfo{year}{2019}).

\bibitem{van2001Quantum}
\bibinfo{author}{van Enk, S.~J.} \& \bibinfo{author}{Fuchs, C.~A.}
\newblock \bibinfo{title}{Quantum state of an ideal propagating laser field}.
\newblock \emph{\bibinfo{journal}{Phys. Rev. Lett.}} \textbf{\bibinfo{volume}{88}}, \bibinfo{pages}{027902} (\bibinfo{year}{2001}).

\bibitem{shannon1949communication}
\bibinfo{author}{Shannon, C.~E.}
\newblock \bibinfo{title}{Communication theory of secrecy systems}.
\newblock \emph{\bibinfo{journal}{Bell Syst. Tech. J.}} \textbf{\bibinfo{volume}{28}}, \bibinfo{pages}{656--715} (\bibinfo{year}{1949}).

\bibitem{wegman1981new}
\bibinfo{author}{Wegman, M.~N.} \& \bibinfo{author}{Carter, J.~L.}
\newblock \bibinfo{title}{New hash functions and their use in authentication and set equality}.
\newblock \emph{\bibinfo{journal}{J. Comput. Syst. Sci.}} \textbf{\bibinfo{volume}{22}}, \bibinfo{pages}{265--279} (\bibinfo{year}{1981}).

\bibitem{krawczyk1994lfsr}
\bibinfo{author}{Krawczyk, H.}
\newblock \bibinfo{title}{Lfsr-based hashing and authentication}.
\newblock In \emph{\bibinfo{booktitle}{Annual International Cryptology Conference}}, \bibinfo{pages}{129--139} (\bibinfo{organization}{Springer}, \bibinfo{year}{1994}).

\bibitem{bennett1995generalized}
\bibinfo{author}{Bennett, C.~H.}, \bibinfo{author}{Brassard, G.}, \bibinfo{author}{Cr{\'e}peau, C.} \& \bibinfo{author}{Maurer, U.~M.}
\newblock \bibinfo{title}{Generalized privacy amplification}.
\newblock \emph{\bibinfo{journal}{IEEE Transactions on Information theory}} \textbf{\bibinfo{volume}{41}}, \bibinfo{pages}{1915--1923} (\bibinfo{year}{1995}).

\bibitem{buhrman2001quantum}
\bibinfo{author}{Buhrman, H.}, \bibinfo{author}{Cleve, R.}, \bibinfo{author}{Watrous, J.} \& \bibinfo{author}{De~Wolf, R.}
\newblock \bibinfo{title}{Quantum fingerprinting}.
\newblock \emph{\bibinfo{journal}{Phys. Rev. Lett.}} \textbf{\bibinfo{volume}{87}}, \bibinfo{pages}{167902} (\bibinfo{year}{2001}).

\bibitem{Pegg:1989:Phase}
\bibinfo{author}{Pegg, D.~T.} \& \bibinfo{author}{Barnett, S.~M.}
\newblock \bibinfo{title}{Phase properties of the quantized single-mode electromagnetic field}.
\newblock \emph{\bibinfo{journal}{Phys. Rev. A}} \textbf{\bibinfo{volume}{39}}, \bibinfo{pages}{1665--1675} (\bibinfo{year}{1989}).

\bibitem{Buzek:1992:Coherent}
\bibinfo{author}{Bu\ifmmode~\check{z}\else \v{z}\fi{}ek, V.}, \bibinfo{author}{Wilson-Gordon, A.~D.}, \bibinfo{author}{Knight, P.~L.} \& \bibinfo{author}{Lai, W.~K.}
\newblock \bibinfo{title}{Coherent states in a finite-dimensional basis: Their phase properties and relationship to coherent states of light}.
\newblock \emph{\bibinfo{journal}{Phys. Rev. A}} \textbf{\bibinfo{volume}{45}}, \bibinfo{pages}{8079--8094} (\bibinfo{year}{1992}).

\bibitem{cao2015discrete}
\bibinfo{author}{Cao, Z.}, \bibinfo{author}{Zhang, Z.}, \bibinfo{author}{Lo, H.-K.} \& \bibinfo{author}{Ma, X.}
\newblock \bibinfo{title}{Discrete-phase-randomized coherent state source and its application in quantum key distribution}.
\newblock \emph{\bibinfo{journal}{New J. Phy.}} \textbf{\bibinfo{volume}{17}}, \bibinfo{pages}{053014} (\bibinfo{year}{2015}).

\bibitem{shao2023phase}
\bibinfo{author}{Shao, S.-F.} \emph{et~al.}
\newblock \bibinfo{title}{Phase-matching quantum key distribution without intensity modulation}.
\newblock \emph{\bibinfo{journal}{Phys. Rev. Appl.}} \textbf{\bibinfo{volume}{20}}, \bibinfo{pages}{024046} (\bibinfo{year}{2023}).

\bibitem{chefles1998optimum}
\bibinfo{author}{Chefles, A.} \& \bibinfo{author}{Barnett, S.~M.}
\newblock \bibinfo{title}{Optimum unambiguous discrimination between linearly independent symmetric states}.
\newblock \emph{\bibinfo{journal}{Phys. Lett. A}} \textbf{\bibinfo{volume}{250}}, \bibinfo{pages}{223--229} (\bibinfo{year}{1998}).

\bibitem{van2002unambiguous}
\bibinfo{author}{Van~Enk, S.}
\newblock \bibinfo{title}{Unambiguous state discrimination of coherent states with linear optics: Application to quantum cryptography}.
\newblock \emph{\bibinfo{journal}{Phys. Rev. A}} \textbf{\bibinfo{volume}{66}}, \bibinfo{pages}{042313} (\bibinfo{year}{2002}).

\bibitem{barnett2009quantum}
\bibinfo{author}{Barnett, S.~M.} \& \bibinfo{author}{Croke, S.}
\newblock \bibinfo{title}{Quantum state discrimination}.
\newblock \emph{\bibinfo{journal}{Adv. Opt. Photonics}} \textbf{\bibinfo{volume}{1}}, \bibinfo{pages}{238--278} (\bibinfo{year}{2009}).

\bibitem{wallden2014minimum}
\bibinfo{author}{Wallden, P.}, \bibinfo{author}{Dunjko, V.} \& \bibinfo{author}{Andersson, E.}
\newblock \bibinfo{title}{Minimum-cost quantum measurements for quantum information}.
\newblock \emph{\bibinfo{journal}{J. Phys. A: Math. and Theor.}} \textbf{\bibinfo{volume}{47}}, \bibinfo{pages}{125303} (\bibinfo{year}{2014}).

\end{thebibliography}

\end{document}


\title{Supplementary Information for ``Unconditionally secure key distribution without quantum channel"}

\author{Hua-Lei Yin}
\email{hlyin@ruc.edu.cn}
\affiliation{Department of Physics and Beijing Key Laboratory of Opto-electronic Functional Materials and Micro-nano Devices, Key Laboratory of Quantum State Construction and Manipulation (Ministry of Education), Renmin University of China, Beijing 100872, China}

\date{\today}

\maketitle

\section{I. Review of quantum measurements}

No quantum measurement can perfectly discriminate between non-orthogonal quantum states. Here, we first provide a brief review of two highly effective quantum measurements for $m$ symmetric coherent states. The first is the optimal unambiguous state discrimination (USD) measurement~\cite{chefles1998optimum,barnett2009quantum}, and the second is the minimum error discrimination measurement~\cite{barnett2009quantum,wallden2014minimum}.

\subsection{A. Unambiguous state discrimination}

For $N$ linearly independent pure state $\{\ket{\psi_{j}}\}$, $j=0,1,\ldots,N-1$, there is a positive operator-valued measure (POVM) $\{\hat{E}_{j}\}_{j=0}^{N}$ with $N+1$ elements that can realize unambiguous state discrimination measurement. These POVM elements can be given by~\cite{chefles1998optimum}
\begin{equation}
\begin{cases}
\hat{E}_{j}&=\frac{P_{j}}{|\langle\psi_{j}|\psi_{j}^{\perp}\rangle|^{2}}\ket{\psi_{j}^{\perp}}\bra{\psi_{j}^{\perp}}, ~~~\forall j=0,1,\ldots,N-1,\\
\\
\hat{E}_{N}&=\hat{E}_{?}=\hat{\textbf{I}}-\sum_{j=0}^{N-1}\hat{E}_{j},\\
\end{cases}
\end{equation}
where $\hat{\textbf{I}}$ is the unit operator and $\hat{E}_{j}\geq 0$ is the positive operator. Operator $\hat{E}_{?}$ leads to a failure of the state discrimination measurement.
Thereinto, for all $i,j=0,1,\ldots,N-1$, we have
\begin{equation}
\begin{cases}
\langle\psi_{i}|\hat{E}_{j}|\psi_{i}\rangle&=P_{i}\delta_{ij},\\
\\
\langle\psi_{i}|\psi_{j}^{\bot}\rangle&=\langle\psi_{j}|\psi_{j}^{\bot}\rangle\delta_{ij},\\
\end{cases}
\end{equation}
where $\delta_{ij}=1$ if $i=j$, otherwise, $\delta_{ij}=0$. The result $j$ is obtained only if the state is $\ket{\psi_{j}}$ and the probability of occurrence is $0\leq P_{j}\leq1$.
The state $\ket{\psi_{j}^{\bot}}$ is orthogonal to all allowed states except for the state $\ket{\psi_{j}}$.

Considering that the prior probability of state $\ket{\psi_{j}}$ is $p_{j}$, thus the density matrix of the system is $\hat{\rho}=\sum_{j=0}^{N-1}p_{j}\ket{\psi_{j}}\bra{\psi_{j}}$.
The USD probability $P_{\rm USD}^{(N)}$ is defined as the total probability of correctly identifying the state,
\begin{equation}
\begin{aligned}
P_{\rm USD}^{(N)}&=\sum_{j=0}^{N-1}{\rm tr}(\hat{E}_{j}\hat{\rho})=\sum_{j=0}^{N-1}p_{j}\langle\psi_{j}|E_{j}|\psi_{j}\rangle\\
&=\sum_{j=0}^{N-1}p_{j}P_{j}.
\end{aligned}
\end{equation}
One can implement unambiguous state discrimination measurements since $m$ symmetric coherent states are linearly independent quantum states.
For $N$ symmetric coherent states $\left\{\ket{\sqrt{\mu}e^{\textbf{i}2\pi j/N}}\right\}_{j=0}^{N-1}$ with a uniform priori probability $p_{j}=1/N$, the optimal USD probability can be written as~\cite{chefles1998optimum}
\begin{equation}
\begin{aligned}\label{eqs1}
P_{\rm USD}^{(N)}=\min_{r=0,1,\ldots,N-1}\sum_{j=0}^{N-1}e^{-\textbf{i}2\pi jr/N}e^{\mu\left(e^{\textbf{i}2\pi j/N}-1\right)}.
\end{aligned}
\end{equation}
Considering the special case $N=2$, the optimal probability $P_{\rm USD}^{(2)}=\min\{1+e^{-2\mu},1-e^{-2\mu}\}=1-e^{-2\mu}$, which is identical to the analytical two pure-state conclusion~\cite{tang2013source} $P_{\rm USD}^{(2)}=1-|\langle-\sqrt{\mu}|\sqrt{\mu}\rangle|=1-e^{-2\mu}$. Note that Eq.~\eqref{eqs1} usually does not yield analytical results; we can only calculate them numerically. However, for small intensity $\mu$, we can use an approximate analytical formula~\cite{van2002unambiguous}
\begin{equation}
\begin{aligned}
P_{\rm USD}^{(N)}&=\min_{r=0,1,\ldots,N-1}\sum_{j=0}^{N-1}e^{-\textbf{i}2\pi jr/N}e^{\mu\left(e^{\textbf{i}2\pi j/N}-1\right)}\\
&\approx\frac{N\mu^{N-1}}{(N-1)!}.
\end{aligned}
\end{equation}

\subsection{B. Minimum error discrimination}

For $N$ possible states $\{\hat{\rho}_{j}\}_{j=0}^{N-1}$ with associated a priori probabilities $\{p_{j}\}_{j=0}^{N-1}$, there is a POVM $\{E_{j}\}_{j=0}^{N-1}$ with $N$ elements that can achieve minimum error discrimination. The sufficient and necessary conditions of this POVM can be written as~\cite{barnett2009quantum}
\begin{equation}
\begin{cases}
\sum_{i=0}^{N-1}p_{i}\hat{\rho}_{i}\hat{E}_{i}-p_{j}\hat{\rho}_{j}&\geq0,~~~\forall j=0,1,\ldots,N-1,\\
\\
\hat{E}_{i}(p_{i}\hat{\rho}_{i}-p_{j}\hat{\rho}_{j})\hat{E}_{j}&=0, ~~~\forall i,j=0,1,\ldots,N-1.\\
\end{cases}
\end{equation}
Note that the two conditions are not independent; i.e., the second condition may be derived from the first condition.
The minimum error discrimination probability is defined as
\begin{equation}
\begin{aligned}
P_{\rm min}=\sum_{j=0}^{N-1}p_{j}\sum_{i\neq j}{\rm tr}(\hat{\rho}_{j}\hat{E}_{i}).
\end{aligned}
\end{equation}

For many cases, the minimum error discrimination measurement is the square-root measurement. In addition, there is an important conclusion for square-root measurements; i.e., for any set of pure states, there is at least one set of prior probabilities such that the minimum error discrimination measurement for this set of states is the square root measurement.
The POVM elements of the square-root measurement can be given by~\cite{barnett2009quantum}
\begin{equation}
\begin{aligned}
\hat{E}_{j}=p_{j}\hat{\rho}^{-1/2}\hat{\rho}_{j}\hat{\rho}^{-1/2},
\end{aligned}
\end{equation}
where $\hat{\rho}=\sum_{j=0}^{N-1}p_{j}\hat{\rho}_{j}$. Obviously, the above operator $E_{j}$ is positive, and $\sum_{j=0}^{N-1}\hat{E}_{j}=\hat{\textbf{I}}$. The square-root measurement is the minimum error discrimination measurement for symmetric coherent states. Considering $N$ symmetric coherent states $\left\{\ket{\psi_{j}}=\ket{\sqrt{\mu}e^{\textbf{i}2\pi j/N}}\right\}_{j=0}^{N-1}$ with a uniform priori probability of $p_{j}=1/N$. The Gram matrix of the states we are trying to distinguish between is an $N\times N$ matrix, where the matrix element $G_{i,j}$ of the $i$-th row and $j$-th column can be defined as
\begin{equation}
\begin{aligned}
G_{i,j}&=\langle\psi_{i}|\psi_{j}\rangle=\langle\sqrt{\mu}e^{\textbf{i}2\pi i/N}|\sqrt{\mu}e^{\textbf{i}2\pi j/N}\rangle\\
&=e^{-\mu\left[1-e^{\textbf{i}2\pi(j-i)/N}\right]},
\end{aligned}
\end{equation}
where $i,j=0,1,\ldots,N-1$. Note that we let the matrix start with zero rows and zero columns instead of one row and one column for consistency. Obviously, the Gram matrix $G$ is a circulant matrix since it relies only on the difference $j-i$. It can be diagonalized with the unitary discrete Fourier transform. The eigenvalue $\lambda_{r}$ of Gram matrix $G$ can be given by
\begin{equation}
\begin{aligned}
\lambda_{r}=\sum_{k=0}^{N-1}c_{k}\omega^{kr}
\end{aligned}
\end{equation}
where we have $r=0,1,\ldots,N-1$, $\omega=e^{\textbf{i}2\pi/N}$ and $c_{k}=e^{-\mu\left(1-e^{\textbf{i}2\pi k/N}\right)}$. The optimal minimum error discrimination probability $P_{\rm min}$ can be written as~\cite{wallden2014minimum}
\begin{equation}
\begin{aligned}
P_{\rm min}&=1-\frac{1}{N^{2}}\left|\sum_{r=0}^{N-1}\sqrt{\lambda_{r}}\right|^{2}\\
&=1-\frac{1}{N^{2}}\left|\sum_{r=0}^{N-1}\sqrt{\sum_{k=0}^{N-1}e^{-\mu(1-e^{\textbf{i}2\pi k/N})+\textbf{i}2\pi kr/N}}\right|^{2}.
\end{aligned}
\end{equation}

\section{II. Proof of the first observation}

\emph{First observation}:
The process of generating the discrete phase-randomized weak coherent state through a random mapping rule is the provable quantum one-way function if the phase number is sufficiently large and the optical intensity remains low.

Each weak coherent state $\ket{\sqrt{\mu}e^{\textbf{i}\theta}}$ is actively generated through phase modulation, where the electrical signal is determined by the bit substring $\vec{x}$. According to the definition of the provable quantum one-way function provided in the main text, an attacker Eve cannot obtain any information about $\vec{x}$ if Eve only have access to the quantum state. This quantum system will be considered as a mixture state rather than the pure state due to the absence of phase information.

To demonstrate the validity of this assertion, we will show that the density matrix of the discrete phase-randomized weak coherent state is equivalent to a \emph{pseudo} photon number mixture state. This \emph{pseudo} photon number mixture state is $\epsilon$-close ($\epsilon$ being an exponentially small number close to zero) to the Poisson-distributed photon number mixture state, which is uncorrelated with the global phase $\theta$. The Poisson-distributed photon number mixture state is identical to the continuous phase-randomized coherent state. Therefore, unambiguous state discrimination cannot be utilized, and each state can only be guessed randomly.

Furthermore, using the random mapping rule, each phase interval (e.g., $[0, \pi)$) also corresponds to random bit substrings, with each bit of the substrings having a $50\%$ probability of being zero. Thus, Eve cannot derive any information about $\vec{x}$ even through phase interval estimation via quantum measurement. It is important to note that $\vec{x}$ must be a true random number generated by quantum or other physical random number generation methods. Consequently, each phase of the weak coherent state is independent and random.

\subsection{A. Random mapping rule}

For $m$ global phases $\theta\in\{0,2\pi/m,4\pi/m,\ldots, 2\pi(m-1)/m\}$, there are $m!$ possible mapping rules. In each probability key distribution (PKD) session, the mapping rule that links the global phase $\theta$ to the bit substring $\vec{x}$ is random and determined by a truly random bit string $\vec{C}$.
Various methods can generate these random mapping rules. Here, we describe a straightforward, albeit non-optimal, method involving true random numbers. Traditionally, true random number resources have been relatively accessible. Recent advancements in quantum random number generation have made it possible to easily generate tens of gigabits per second of quantum random numbers.

\begin{figure*}[t]
\centering
\includegraphics[width=18cm]{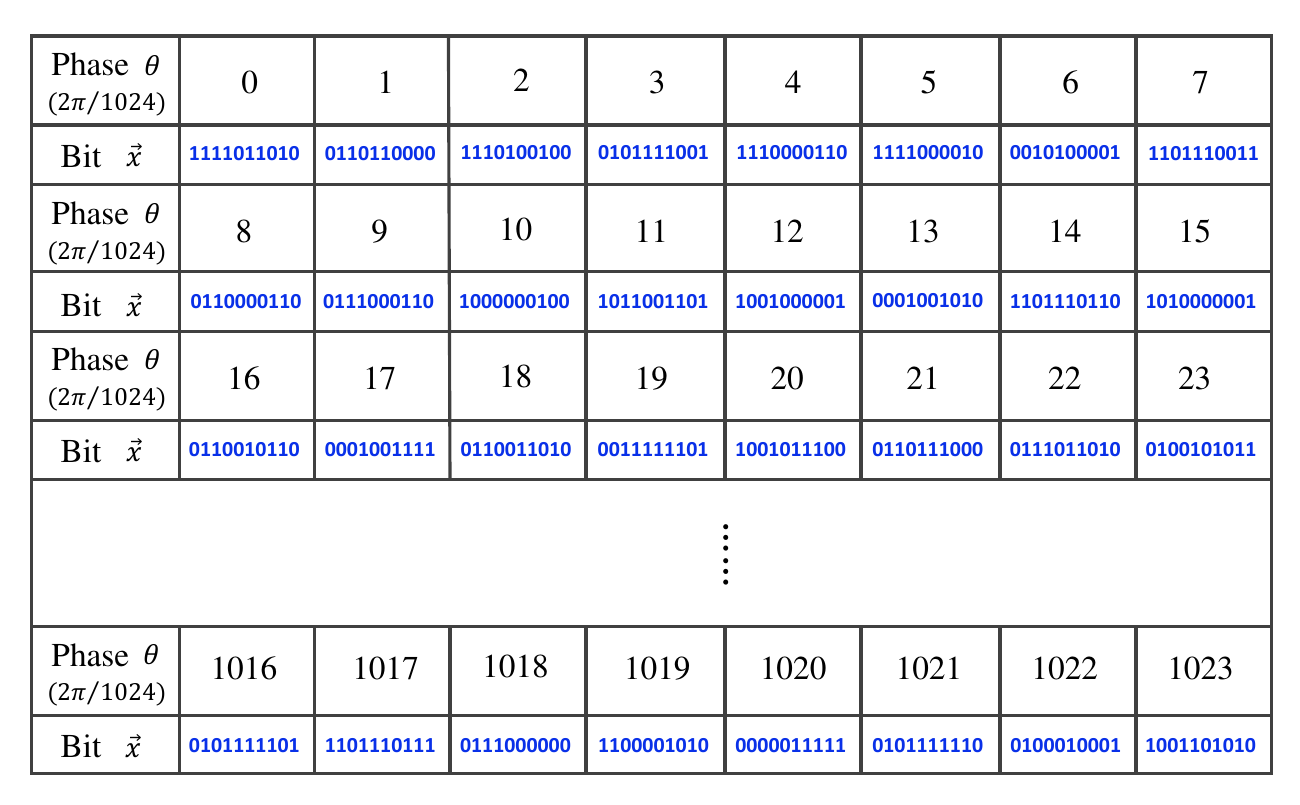}
\caption{The random mapping rule of some one PKD session. The basic unit of the phase row is $2\pi/1024$; thus, 1023 denotes that the phase $\theta=2\pi/1024\times1023$. The bit substring $\vec{x}=1111011010=\vec{c}_{0}$ is the first to appear, so it corresponds to phase $\theta=0$. The bit substring $\vec{x}=1001101010=\vec{c}_{1023}$ is the last to appear, so it corresponds to the phase $\theta=2\pi/1024\times1023$.
} \label{sf1}
\end{figure*}

As illustrated in Fig.~\ref{sf1}, consider one user, such as Alice, who sorts a total of $m$ phases, ranging from 0 to $2\pi(m-1)/m$, into a sequence ordered from smallest to largest. Each phase corresponds to a $\log_{2}m$-bit value. Given a true random number string $\vec{B}$ with a sufficiently large number of bits, $O(10m\log_{2}m)$, Alice treats each $\log_{2}m$-bit of $\vec{B}$ segment as a bit substring. The bit substring is then filled into $\vec{C}$ according to the order in which each $\log_{2}m$-bit substring first appears in $\vec{B}$. For instance, if $m=1024$, then $\log_{2}m=10$, and the random bit string $\vec{B}\in\{0,1\}^{10^{5}}$  is divided into $10^4$ bit substrings. Suppose the first six bit substrings are $\vec{B}_{1}=1111011010$, $\vec{B}_{2}=0110110000$, $\vec{B}_{3}=1110100100$, $\vec{B}_{4}=0101111001$, $\vec{B}_{5}=0110110000$ and $\vec{B}_{6}=1110000110$. In this case, $\vec{B}_{5}$ is discarded since it is a duplicate of $\vec{B}_{2}$, which already appeared earlier. Following this method, the bit string that determines the random mapping rule is $\vec{C}=\vec{c}_{0}||\vec{c}_{1}||\vec{c}_{2}||\cdots||\vec{c}_{m-1}$, where $\vec{c}_{0}=\vec{B}_{1}=1111011010$, $\vec{c}_{1}=\vec{B}_{2}=0110110000$, $\vec{c}_{2}=\vec{B}_{3}=1110100100$, $\vec{c}_{3}=\vec{B}_{4}=0101111001$, $\vec{c}_{4}=\vec{B}_{6}=1110000110$, and so forth. Consequently, the bit substring $\vec{x}=1111011010=\vec{c}_{0}$  (which is 968 in decimal notation) corresponds to global phase $\theta=0$, the bit substring $\vec{x}=0110110000=\vec{c}_{1}$ corresponds to phase $\theta=2\pi/1024\times 1$, the bit substring $\vec{x}=1110100100=\vec{c}_{2}$ corresponds to global phase $\theta=2\pi/1024\times2$, the bit substring $\vec{x}=0101111001=\vec{c}_{3}$ corresponds to global phase $\theta=2\pi/1024\times3$, and the bit substring $\vec{x}=1110000110=\vec{c}_{4}$ corresponds to global  phase $\theta=2\pi/1024\times4$. The bit substring $\vec{x}=\vec{c}_{j}$ corresponds to global phase $\theta=2\pi/1024\times j$. The detailed mapping rule for a particular session is shown in Fig.~\ref{sf1}.
In each PKD session, $m\log_{2}m$ bits (which is greater than $\log_{2}(m!)$) are used to determine a specific random mapping rule, and these rules vary with each session.

The specific random mapping rule $\vec{C}$ in each session is known only to Alice. To share this rule with Bob, Alice must encrypt it and transmit it using a one-time pad, which requires an $m\log_{2}m$-bit pre-shared secret key. Additionally, to ensure that the mapping rules have not been tampered with by Eve, it is necessary to use information-theoretically secure authentication methods~\cite{wegman1981new,krawczyk1994lfsr}. To reduce the amount of secret key required for message authentication, Alice and Bob can authenticate multiple mapping rules from several PKD sessions simultaneously. Since the random mapping rule is determined by true random numbers, all possible mapping rules occur with equal probability. Consequently, from the view of Eve, each global phase $\theta$ is uniformly and randomly associated with all possible $\log_{2}m$-bit substrings.

\subsection{B. Discrete phase-randomized coherent state}

We now quantitatively analyze the relationship between discrete and continuous phase-randomized coherent states. First, we demonstrate that the continuous phase-randomized coherent state $\ket{\sqrt{\mu}e^{\textbf{i}\theta}}$ can be considered uncorrelated with the global phase $\theta$. For the discrete phase-randomized coherent state $\ket{\sqrt{\mu}e^{\textbf{i}\theta}}$, where $\theta\in\{0,2\pi/m,4\pi/m,\ldots, 2\pi(m-1)/m\}$, the discrete case approaches the continuous case as $m\rightarrow\infty$. Specifically, the discrete phase values $\theta$ will cover the continuous range $[0, 2\pi)$ in the limit of large $m$.
The density matrix of the continuous phase-randomized weak coherent state can be given by
\begin{equation}
\begin{aligned}\label{eqs2}
\hat{\rho}^{\rm con}=\frac{1}{2\pi}\int_{0}^{2\pi}\ket{\sqrt{\mu}e^{\textbf{i}\theta}}\bra{\sqrt{\mu}e^{\textbf{i}\theta}}d\theta=\sum_{k=0}^{\infty}e^{-\mu}\frac{\mu^{k}}{k! }\ket{k}\bra{k}=\sum_{k=0}^{\infty}P^{\mu}(k)\ket{k}\bra{k},
\end{aligned}
\end{equation}
where $P^{\mu}(k)=e^{-\mu}\mu^{k}/k!$ represents the Poisson distribution for the Fock state \(\ket{k}\), and the right-hand side of Eq.~\eqref{eqs2} is a mixture of these Poisson-distributed Fock states. The density matrix is identical in both cases, making them indistinguishable. Clearly, the global phase of a Fock state has no physical significance; the statistics of measurements predicted for $e^{\textbf{i}\theta}\ket{k}$ and $\ket{k}$ are the same. Thus, from an observational perspective, the states $e^{\textbf{i}\theta}\ket{k}$ and $\ket{k}$ are equivalent. Consequently, for Eve, the global phase is independent of the continuous phase-randomized weak coherent state. In other words, without knowledge of the phase information, the continuous phase-randomized quantum weak coherent state appears as a photon number mixture state. For Eve, this means the global phase is entirely random, and no useful information can be extracted from it. This is what we refer to as the source of rigorous one-wayness. Conversely, the preparator perceives the system as a pure state rather than a photon number mixture state, given that they generated it.

To show this more intuitively, we introduce the normalized quantum phase operator~\cite{Pegg:1989:Phase,Buzek:1992:Coherent}
\begin{equation}
\begin{aligned}
e^{\textbf{i}\hat{\phi}_{\theta}}=\lim_{l\rightarrow\infty}\sum_{j=0}^{l}e^{\textbf{i}\theta_{j}}\ket{\theta_{j}}\bra{\theta_{j}},
\end{aligned}
\end{equation}
where $j=0,1,\ldots,l$ and phase state $\ket{\theta_{j}}$ is the eigenstate of phase operator $\hat{\phi}_{\theta}$, since we can prove that $e^{\textbf{i}\hat{\phi}_{\theta}}\ket{\theta_{j}}=e^{\textbf{i}\theta_{j}}\ket{\theta_{j}}$ and $\hat{\phi}_{\theta}=\lim_{l\rightarrow\infty}\sum_{j=0}^{l}\theta_{j}\ket{\theta_{j}}\bra{\theta_{j}}$.
The phase state $\ket{\theta_{j}}$ is defined in $l+1$ dimension Fock state space
\begin{equation}
\begin{aligned}
\ket{\theta_{j}}=\lim_{l\rightarrow\infty}\frac{1}{\sqrt{l+1}}\sum_{k=0}^{l}e^{\textbf{i}k\theta_{j}}\ket{k},
\end{aligned}
\end{equation}
where phase $\theta_{j}=\theta_{0}+2\pi j/(l+1)$ and $\theta_{0}$ is the reference state. The phase state satisfies the complete orthogonal condition $\langle\theta_{j}|\theta_{h}\rangle=\delta_{jh}$ and $\hat{\textbf{I}}_{l+1}=\sum_{j=0}^{l}\ket{\theta_{j}}\bra{\theta_{j}}$. Therefore, for $k\leq l$, one can expand the Fock state $\ket{k}$ with a phase state
\begin{equation}
\begin{aligned}
\ket{k}=\lim_{l\rightarrow\infty}\frac{1}{\sqrt{l+1}}\sum_{j=0}^{l}e^{-\textbf{i}k\theta_{j}}\ket{\theta_{j}},
\end{aligned}
\end{equation}
which means that each Fock state $\ket{k}$ is a uniform superposition of phase states $\theta_{j}$. Therefore, one obtained phase via any quantum measurement will be completely random given the mixture of the Fock state.

For $m$ discrete global phases, the density matrix of the discrete phase-randomized weak coherent state can be written as~\cite{cao2015discrete,shao2023phase}
\begin{equation}
\begin{aligned}
\hat{\rho}^{\rm dis}(m)=\frac{1}{m}\sum_{j=0}^{m-1}\ket{\sqrt{\mu}e^{\textbf{i}\frac{2\pi}{m}j}}\bra{\sqrt{\mu}e^{\textbf{i}\frac{2\pi}{m}j}} = \sum_{k=0}^{m-1} P^{\mu}_{m}(k)\ket{\lambda_{k}}\bra{\lambda_{k}},
\end{aligned}
\end{equation}
which can be regarded as the mixture of pseudo photon-number states $\ket{\lambda_{k}}$ with probability $P^{\mu}_{m}(k)$.
The probability $P^{\mu}_{m}(k)$ and pseudo photon-number state $\ket{\lambda_{k}}$ can be given by
\begin{equation}
\begin{aligned}
P^{\mu}_{m}(k)&=e^{-\mu}\sum_{l=0}^{\infty}\frac{\mu^{lm+k}}{(lm+k)! },\\
\ket{\lambda_{k}}&=\frac{e^{-\mu/2}}{\sqrt{P^{\mu}_{m}(k)}}\sum_{l=0}^{\infty}\frac{(\sqrt{\mu})^{lm+k}}{\sqrt{(lm+k)!}}\ket{lm+k}.
\end{aligned}
\end{equation}
Obviously, the probability will become the Poisson distribution, and state $\lambda_{k}$ will become the Fock state if $m\rightarrow\infty$, i.e., $\lim_{m\rightarrow\infty}P^{\mu}_{m}(k)\equiv e^{-\mu}\mu^{k}/k!$ and $\lim_{m\rightarrow\infty}\ket{\lambda_{k}}\equiv\ket{k}$. In the asymptotic limit case, we have $m=2\pi/d\theta$; thus,
\begin{equation}
\begin{aligned}
\lim_{m\rightarrow\infty}\frac{1}{m}\sum_{j=0}^{m-1}\ket{\sqrt{\mu}e^{\textbf{i}\frac{2\pi}{m}j}}\bra{\sqrt{\mu}e^{\textbf{i}\frac{2\pi}{m}j}} = \frac{1}{2\pi}\int_{0}^{2\pi}\ket{\sqrt{\mu}e^{\textbf{i}\theta}}\bra{\sqrt{\mu}e^{\textbf{i}\theta}}d\theta.
\end{aligned}
\end{equation}

For finite $m$, we can calculate the probability difference $\Delta(\mu,m,k)$ and trace distance $D(\ket{\lambda_{k}},\ket{k})$ between the Fock state and the pseudo photon-number state. For $0\leq k\leq m-1$, we have the probability difference
\begin{equation}
\begin{aligned}
\Delta(\mu,m,k)&=\frac{P^{\mu}_{m}(k)-P^{\mu}(k)}{P^{\mu}(k)}=\sum_{l=1}^{\infty}\frac{\mu^{lm}k!}{(lm+k)! }.
\end{aligned}
\end{equation}
For $0\leq k\leq m-1$, we have the trace distance
\begin{equation}
\begin{aligned}
D(\ket{\lambda_{k}},\ket{k})&=\frac{1}{2}{\rm tr}|\ket{\lambda_{k}}\bra{\lambda_{k}}-\ket{k}\bra{k}|=\sqrt{1-F(\ket{\lambda_{k}},\ket{k})^{2}}\\
&=\sqrt{1-\frac{1}{1+\sum_{l=1}^{\infty}\frac{\mu^{lm}k!}{(lm+k)! }}}=\sqrt{1-\frac{1}{1+\Delta(\mu,m,k)}}\\
&<\sqrt{\Delta(\mu,m,k)}=\sqrt{\sum_{l=1}^{\infty}\frac{\mu^{lm}k!}{(lm+k)! }}.
\end{aligned}
\end{equation}
Obviously, for $\Delta(\mu,m,k)\leq\Delta(\mu,m,0)$, we have $\Delta(0.1,1024,0)=1.85\times10^{-3664}$.

For the phase-randomized coherent state, we consider that the purification of this system
\begin{equation}
\begin{aligned}
\ket{\psi}_{AE}=\frac{1}{\sqrt{m}}\sum_{j=0}^{m-1}\ket{j}_{A}\ket{\sqrt{\mu}e^{\textbf{i}\theta_{j}}}_{E},
\end{aligned}
\end{equation}
where $\theta_{j}=2\pi j/m$, only Alice owns the qudit system $A$, and Eve can access the optical mode $E$. Deterministic phase information $\theta_{j}$ can be obtained by measuring qudit $\ket{j}_{A}$, i.e., from the view of Alice, the optical mode will become the pure state $\ket{\sqrt{\mu}e^{\textbf{i}\theta_{j}}}_{E}$. However, Eve does not have access to the qudit system. From the view of Eve, the optical mode becomes the mixed state
\begin{equation}
\begin{aligned}
\hat{\rho}_{E}^{\rm dis}(m)={\rm tr}_{A}\ket{\psi}_{AE}\bra{\psi}_{AE}=\frac{1}{m}\sum_{j=0}^{m-1}\ket{\sqrt{\mu}e^{\textbf{i}\frac{2\pi}{m}j}}_{E}\bra{\sqrt{\mu}e^{\textbf{i}\frac{2\pi}{m}j}}.
\end{aligned}
\end{equation}
Since $\lim_{m\rightarrow\infty}\hat{\rho}_{E}^{\rm dis}(m)=\hat{\rho}_{E}^{\rm con}=\sum_{k=0}^{\infty}e^{-\mu}\mu^{k}/k!\times\ket{k}_{E}\bra{k}$, the system of the Eve is completely decoupled from the phase information $\theta_{j}$ of Alice if $m\rightarrow\infty$. In other words, Eve can only randomly guess the phase if he or she has $\hat{\rho}_{E}^{\rm con}$. However, $\hat{\rho}_{E}^{\rm dis}(m)\neq\hat{\rho}_{E}^{\rm con}$ for finite $m$, which leads Eve to obtain some phase information. For example, for $m=2$, the probability of obtaining an optimum unambiguous state discrimination measurement is $P_{\rm USD}^{(2)}=1-e^{-2\mu}$. Fortunately, the amount of leaked to Eve's information decreases exponentially with increasing $m$.

To define the secrecy of a phase, we consider the quantum state $\hat{\rho}_{E}$, which describes the correlation between Alice's phase information and the eavesdropper Eve (for any given attack strategy). The phase is called $\epsilon$-secret from Eve if it is $\epsilon$-close to a uniformly distributed phase that is uncorrelated with the eavesdropper, that is, if
\begin{equation}
\begin{aligned}
\frac{1}{2}\left\|\hat{\rho}_{E}-\hat{\rho}_{E}^{\rm ideal}\right\|_{1}\leq\epsilon,
\end{aligned}
\end{equation}
where $\|\cdot\|_{1}$ is the trace norm and $\hat{\rho}_{E}^{\rm ideal}$ is the ideal state that is completely decoupled from Alice's phase. According to this definition, the quantum state is $\hat{\rho}_{E}=\hat{\rho}_{E}^{\rm dis}(m)$ for $m$ symmetric coherent states, and the ideal state is $\hat{\rho}_{E}^{\rm ideal}=\hat{\rho}_{E}^{\rm con}$. Let $\lambda_{n}$ be the $n$-th eigenvalue of $\hat{\rho}_{E}^{\rm dis}(m)-\hat{\rho}_{E}^{\rm con}$; the trace distance can be given by
\begin{equation}
\begin{aligned}
D\left(\hat{\rho}_{E}^{\rm dis}(m),\hat{\rho}_{E}^{\rm con}\right)&=\frac{1}{2}\left\|\hat{\rho}_{E}^{\rm dis}(m)-\hat{\rho}_{E}^{\rm con}\right\|_{1}\\
&=\frac{1}{2}{\rm tr}\left|\hat{\rho}_{E}^{\rm dis}(m)-\hat{\rho}_{E}^{\rm con}\right|\\
&=\frac{1}{2}\sum_{n=0}^{\infty}|\lambda_{n}|.
\end{aligned}
\end{equation}
The above formula can be calculated numerically only. This process is highly complicated because there is an infinite number of photons. Here, we consider the photon number truncated to $m-1$; the density matrices $\hat{\rho}_{E}^{\rm dis}(m)$ and $\hat{\rho}_{E}^{\rm con}$ are all $m\times m$ diagonal matrices. Thus, we have the approximate result
\begin{equation}
\begin{aligned}
\epsilon&=D\left(\hat{\rho}_{E}^{\rm dis}(m),\hat{\rho}_{E}^{\rm con}\right)=\frac{1}{2}\left|\left|\hat{\rho}_{E}^{\rm dis}(m)-\hat{\rho}_{E}^{\rm con}\right|\right|_{1}\\
&\approx\frac{1}{2}\sum_{k=0}^{m-1}\Delta(\mu,m,k)P^{\mu}(k)\\
&=\frac{e^{-\mu}}{2}\sum_{k=0}^{m-1}\sum_{l=1}^{\infty}\frac{\mu^{lm+k}}{(lm+k)!}\\
&\approx\frac{e^{-\mu}\mu^{m}}{2(m!)}.
\end{aligned}
\end{equation}
where we assume that $m$ is large enough, for example, $m\geq 100$.
If $m=1024$ and $\mu=0.1$, we have $\epsilon=8.35\times10^{-3665}$.

Furthermore, we can use the USD and the minimum error discrimination measurements for analysis from another perspective. If $m=1024$ and $\mu=0.1$, the optimal USD probability and minimum error discrimination measurement probability can be given by
\begin{equation}
\begin{aligned}
P_{\rm USD}^{(m)}&=\frac{m\mu^{m-1}}{(m-1)!}=1.94\times10^{-3657},\\
\end{aligned}
\end{equation}
and
\begin{equation}
\begin{aligned}
P_{\rm min}&=1-\frac{1}{m^{2}}\left|\sum_{r=0}^{m-1}\sqrt{\sum_{k=0}^{m-1}e^{-\mu(1-e^{\textbf{i}2\pi k/m})+\textbf{i}2\pi kr/m}}\right|^{2}=99.83\%.
\end{aligned}
\end{equation}
Obviously, it is impossible for Eve to know with certainty what the phase of one optical pulse is through the USD measurement.
If Eve does not exploit the received quantum state, he can randomly guess the phase, and the error probability of random guessing is $1-1/1024=99.90\%$. Compared with the error probabilities of minimum error discrimination measurement and random guessing, one can find that the quantum states received by Eve offer almost no advantage for large $m$ and small $\mu$.

\section{III. Proof of the second observation}

\emph{Second observation}:
Alice and Bob can share themselves-generated quantum random numbers in many communication rounds with unconditional security via a fixed but long secret key and a per-update but short secret key if these quantum random numbers are not leaked to Eve when they are used.

It is well established that the one-time pad is the only information-theoretically secure encryption system. The one-time pad requires two conditions: the secret key used for encryption must be updated for each use and must be at least as long as the message, and the key must be truly random and kept entirely secret by the communicating parties. Importantly, sharing a true random number rather than the message in the second observation is not an actual encryption process but rather a negotiation of random information. Hence, we remark that the second observation does not contradict the conclusion about the one-time pad.

To further clarify our second observation, we will first review various types of attacks in cryptanalysis~\cite{menezes2018handbook}, explain why the one-time pad is an information-theoretically secure encryption scheme, and contrast it with why a (non-one-time pad) stream cipher does not offer the same level of security.

\subsection{A. Types of attack utilized in cryptanalysis}

The goal of cryptanalysis is to efficiently and fully recover plaintext from ciphertext, and it often involves deducing the decryption key used for past or future messages. The four most important types of attacks are as follows:

1. \emph{Ciphertext-only attack}: Cryptanalysts attempt to deduce the decryption key or plaintext using only the available ciphertext. This is the most challenging scenario for cryptanalysts, as the only information they have is the ciphertext itself.

2. \emph{Known-plaintext attack}: Cryptanalysts try to uncover the decryption key or algorithm by combining intercepted ciphertext with known plaintext-ciphertext pairs. Executing this attack is typically only slightly more challenging than a ciphertext-only attack.

3. \emph{Chosen-plaintext attack}: Cryptanalysts can select specific plaintexts and obtain their corresponding ciphertexts through the encryption system. This type of attack is more potent than a known-plaintext attack but requires additional resources.

4. \emph{Chosen-ciphertext attack}: Cryptanalysts can choose ciphertexts and obtain the corresponding plaintexts. If they have access to decryption equipment (though not necessarily the decryption key), they can use this attack to deduce plaintexts from various ciphertexts without direct access to the decryption key. This attack is the most powerful and is primarily used against public key cryptosystems, though it is traditionally the most challenging to implement.

In addition to the four attacks mentioned, there are other types such as adaptive chosen-plaintext attacks and adaptive chosen-ciphertext attacks. Powerful techniques like differential cryptanalysis and linear cryptanalysis typically belong to known-plaintext attacks. Similarly, a brute-force search attack, even with unlimited computational resources, relies on known plaintext-ciphertext pairs to verify its results, thus also being categorized as a known-plaintext attack.

\subsection{B. One-time pad encryption and perfect secrecy}

Information-theoretical security, also known as unconditional security, ensures that a cryptosystem maintains its security even if the eavesdropper possesses unlimited computational resources. Unconditional security in cryptosystems is referred to as perfect secrecy. To define perfect secrecy, let $\vec{m}$, $\vec{m}_{0}$, and $\vec{m}_{1}$ represent arbitrary plaintexts, and let $\mathcal{M}$ denote the plaintext space. Let $\vec{c}$ be an arbitrary ciphertext, and $\mathcal{C}$ denote the ciphertext space. Let $\vec{k}$ be an arbitrary key, and $\mathcal{K}$ represent the key space. ${\rm Enc}_{\vec{k}}(\vec{m})$ (${\rm Dec}_{\vec{k}}(\vec{m})$) denotes the encryption (decryption) algorithm using key $\vec{k}$ and message $\vec{m}$.

\emph{Perfect secrecy:} for $\forall~\vec{m}_{0}, \vec{m}_{1}\in \mathcal{M}$ ($|\vec{m}_{0}|=|\vec{m}_{1}|$) and $\forall~\vec{c}\in \mathcal{C}$, we call a symmetric encryption system with the property of perfect secrecy if we have
\begin{equation}\label{eqs28}
\begin{aligned}
{\rm Pr}\left[{\rm Enc}_{\vec{k}}(\vec{m}_{0})=\vec{c}\right]={\rm Pr}\left[{\rm Enc}_{\vec{k}}(\vec{m}_{1})=\vec{c}\right],
\end{aligned}
\end{equation}
where $\vec{k}\in\mathcal{K}$ is uniformly random.
The probability of obtaining the ciphertext $\vec{c}$ from the encryption of any two plaintexts within the plaintext space $\mathcal{M}$ is identical. This implies that no information is disclosed by the ciphertext, meaning that an attacker cannot determine which plaintext corresponds to the given ciphertext $\vec{c}$, even if they use unlimited computational resources to enumerate all possible keys in the key space $\mathcal{K}$.

\emph{One-time pad:} For each encryption session, let $\mathcal{M}=\mathcal{C}=\mathcal{K}=\{0,1\}^{n}$, $\vec{m}\in\mathcal{M}$, $\vec{k}\in\mathcal{K}$ and $\vec{c}\in\mathcal{C}$; the encryption and decryption algorithms are
\begin{equation}
\begin{aligned}
{\rm Enc}_{\vec{k}}(\vec{m}): \vec{c}=\vec{m}\oplus \vec{k},\\
{\rm Dec}_{\vec{k}}(\vec{m}): \vec{m}=\vec{c}\oplus \vec{k},
\end{aligned}
\end{equation}
where the random key $\vec{k}$ must be used only once. Obviously, the one-time pad has the property of perfect secrecy since, given $\vec{c}$, for any one $\vec{m}$, there is only one corresponding $\vec{k}$. Eq.~\eqref{eqs28} is naturally satisfied in one-time pad.

\emph{Shannon law}: Let $({\rm Enc}_{\vec{k}}(\vec{m}), {\rm Dnc}_{\vec{k}}(\vec{m}))$ be a symmetric encryption scheme defined on $(\mathcal{K}, \mathcal{M}, \mathcal{C})$; if it is a perfect secrecy, then $|\mathcal{K}|\geq|\mathcal{M}|$. Thus, $|\mathcal{K}|=|\mathcal{M}|$ of one-time pad is the optimal scheme.

We highlight the following points. First, perfect secrecy is concerned with ciphertext-only attacks, meaning the attacker only has access to the eavesdropped ciphertext. Second, since the key is used only once and is updated in each session, the plaintexts, keys, and ciphertexts from other sessions are irrelevant to any single session attack. Consequently, attacks such as known-plaintext attacks are ineffective. Specifically, in a one-time pad scheme, other types of attacks are futile because they provide no useful information to the attacker. A brute-force search attack, even with unlimited computational resources, is also ineffective, as the correctness of the attack results cannot be verified. For instance, if an attacker correctly enumerates a key and obtains the correct plaintext, they cannot distinguish it from other plaintexts. As a result, the attacker is left with no option but to make a completely random guess.

\emph{Stream ciper}: Let $G$ be an efficient computable deterministic function
\begin{equation}
\begin{aligned}
G(\vec{k}):~\vec{k}\in\{0,1\}^{n}\rightarrow \vec{\texttt{k}}_{\rm pse}\in\{0,1\}^{N},~~~n<<N,
\end{aligned}
\end{equation}
where the extended pseudorandom key $\vec{\texttt{k}}_{\rm pse}$ is used in a streaming manner across numerous encryption sessions. The encryption and decryption algorithms also include the XOR operation
\begin{equation}
\begin{aligned}
{\rm Enc}_{\vec{k}}(\vec{m}): \vec{c}=\vec{m}\oplus G(\vec{k}),\\
{\rm Dec}_{\vec{k}}(\vec{m}): \vec{m}=\vec{c}\oplus G(\vec{k}).
\end{aligned}
\end{equation}

Note that in a stream cipher, a short key seed $\vec{k}$ is used for many encryption sessions. According to Shannon's law, a stream cipher cannot achieve perfect secrecy because $|\vec{k}| \ll |\vec{m}|$. An obvious vulnerability is the known-plaintext attack, as the pseudorandom keys $\vec{\texttt{k}}_{\rm pse}$ used in different encryption sessions are not completely independent but related. Consequently, known plaintext-ciphertext pairs can provide useful information to verify the correctness of results in brute-force search attacks. Immunity to known-plaintext attacks is a crucial factor in determining whether a cryptosystem possesses perfect secrecy.

\subsection{C. The security of random information negotiation with $\vec{R}_{n}(\vec{x}_{a})$}

With the above knowledge, we can now easily prove our second observation. Let us first look at the concrete implementation in the main text. Given a fixed but long secret key $\vec{K}_{\rm fix}\in\{0,1\}^{s+t-1}$, a fully random Toeplitz matrix $\textbf{H}_{st}$ with $s\times t$ ($s<<t$) can be constructed by using the secret key $\vec{K}_{\rm fix}=[h_{0},h_{1},\cdots,h_{s+t-1}]$, as follows:
\begin{equation}
\begin{aligned}\label{eqs32}
\textbf{H}_{st}=\left[{\begin{array}{cccccc}
h_{s-1} & h_{s} & h_{s+1} & \cdots & h_{s+t-3}& h_{s+t-2}\\
h_{s-2} & h_{s-1} & h_{s}& \cdots & h_{s+t-4} & h_{s+t-3}\\
h_{s-3} & h_{s-2} & h_{s-1} &\cdots & h_{s+t-5} & h_{s+t-4}\\
\vdots & \vdots & \vdots & \ddots & \vdots & \vdots\\
h_{1} & h_{2} & h_{3} & \cdots & h_{t-1} & h_{t}\\
h_{0} & h_{1} &  h_{2} &\cdots & h_{t-2} &h_{t-1}\\
\end{array}}\right].
\end{aligned}
\end{equation}
Note that $\textbf{H}_{st}$ will be reused in many PKD sessions, for example, $10^{4}$ sessions. For each PKD session, a short but per-update secret key $\vec{K}\in\{0,1\}^{s}$ is exploited to generate the data string $\vec{D}\in\{0,1\}^{t}$,
\begin{equation}
\begin{aligned}\label{eqs33}
\vec{D}=\vec{K}\cdot \textbf{H}_{st}.
\end{aligned}
\end{equation}
Let $\vec{K}=[k_{0},k_{1},\cdots,k_{s-1}]$ and $\vec{D}=[d_{0},d_{1},\cdots,d_{t-1}]$ be the row vector; thus, we have
\begin{equation}
\begin{aligned}\label{eqs34}
d_{j}&=k_{0}h_{s+j-1}\oplus k_{1}h_{s+j-2}\oplus k_{2}h_{s+j-3}\oplus\cdots\oplus k_{s-1}h_{j}\\
&=\sum_{i=0}^{s-1}\oplus k_{i}h_{s+j-i-1},
\end{aligned}
\end{equation}
where $j=0,1,\cdots,t-1$.
Note that in each PKD session, both Alice and Bob share the same secret keys $\vec{K}_{\rm fix}$ and $\vec{K}$, resulting in the identical data string $\vec{D}$. According to Eq.~\eqref{eqs34}, each bit $d_{j}$ of $\vec{D}$ in the first PKD session is independent and random. Consequently, from Eve's perspective, the data string $\vec{D}$ in the first PKD session appears as a random $t$-bit string.
In other words, the value of the bit string $\vec{D}$ can be all $2^{t}$ cases from the bit string $00\cdots0$ to $00\cdots0$ with equal probability.

In the first PKD session, the data string $\vec{D}$ is a random bit string since $\textbf{H}_{st}$ has not been utilized. Let the first $n\log_{2}m$ bits in $\vec{D}$ form bit string $\vec{D}_{n}$, i.e., $\vec{D}_{n}=[d_{0},d_{1},\cdots,d_{n\log_{2}m}]\subseteq \vec{D}$. Let $\vec{R}_{n}(\vec{x}_{a})\in\{0,1\}^{n\log_{2}m}$ represent the random bit string used for discrete phase modulation. Alice shares the random bit string $\vec{R}_{n}(\vec{x}_{a})$ with Bob by sending the XOR result $\vec{c}_{n}=\vec{R}_{n}(\vec{x}_{a})\oplus \vec{D}_{n}$. Bob retrieves the random bit string $\vec{R}_{n}(\vec{x}_{a})$ via the XOR operation $\vec{R}_{n}(\vec{x}_{a})=\vec{c}_{n}\oplus \vec{D}_{n}$. The random bit string $\vec{R}_{n}(\vec{x}_{a})$ is generated solely by Alice and is not the message itself. Although  bit string $\vec{R}_{n}(\vec{x}_{a})$ is also used for discrete phase modulation, the corresponding discrete phase-randomized weak coherent state with a random mapping rule does not leak information to Eve due to the rigorous one-wayness of provable quantum one-way function. Additionally, $\vec{R}_{n}(\vec{x}_{a})$ is used to generated quantum state only once and will be updated in the next PKD session. Therefore, Eve can only exploit ciphertext-only attacks, with no other attack types being viable. In the first PKD session, Eve only has access to the ciphertext $\vec{c}_{n}$ and lacks knowledge of $\vec{R}_{n}(\vec{x}_{a})$ or $\vec{D}_{n}$, which ensures perfect secrecy since $|\vec{R}_{n}(\vec{x}_{a})|$=$|\vec{D}_{n}|$. Consequently, $\textbf{H}_{st}$ remains completely random and non-informative to Eve even after the first PKD session.

In the second PKD session, all procedures are identical to those of the first PKD session, except that the unknown Toeplitz matrix $\textbf{H}_{st}$ is reused.
Note that $\textbf{H}_{st}$ remains entirely random from Eve's perspective in the second PKD session.
Consequently, only ciphertext-only attacks are feasible for Eve, with no other attack methods being effective. Eve has no knowledge of $\vec{R}_{n}(\vec{x}_{a})$ or $\vec{D}_{n}$, ensuring that the sharing of $\vec{R}_{n}(\vec{x}_{a})$ maintains perfect secrecy. Indeed, subsequent PKD sessions follow the same protocol as the second session. Thus, we conclude that Alice and Bob achieve information-theoretical security when sharing the random bit string $\vec{R}_{n}(\vec{x}_{a})$ in all PKD sessions.

It is crucial to emphasize that the primary reason Eve cannot exploit the correlation of $\vec{D}$ across different PKD sessions is that $\vec{R}_{n}(\vec{x}_{a})$ is a true random number rather than a message. Random bit string $\vec{R}_{n}(\vec{x}_{a})$ used to generate quantum system can be proven to maintain rigorous one-wayness (no information is leaked in provable quantum one-way function). Eve does not have access to $\vec{R}_{n}(\vec{x}_{a})$ for each PKD session, meaning she lacks any known plaintext-ciphertext pairs. Consequently, brute-force search attacks using unlimited computational resources are ineffective.
The method for generating the unknown data string $\vec{D}$, as outlined by Eqs.~\eqref{eqs32} and \eqref{eqs33}, is not unique. The generation process must satisfy one condition: the data string $\vec{D}$ must appear completely random to Eve in each session, i.e., $\vec{D}$ must yield all possible bit strings.
It is important to note that our random information negotiation method is not suitable for cryptosystems designed to transmit messages, as it would essentially function as a stream cipher in such cases. Obviously, message may be leaked to Eve. In that scenario, Eve could potentially acquire known plaintext-ciphertext pairs and use brute-force search attacks to verify correctness and try to find the fixed secret key $\vec{K}_{\rm fix}\in\{0,1\}^{s+t-1}$.

Here, to aid in understanding why the repeated use of a single key can still achieve unconditional security, let us recall the unconditionally secure authentication by using random universal$_{2}$ hashing, for example  linear-feedback-shift-register-based (LFSR-based) Toeplitz matrix~\cite{krawczyk1994lfsr}.
The LFSR-based Toeplitz matrix is determined by an irreducible polynomial $p(x)=x_{n}+a_{n-1}x^{n-1}+...+a_{1}x+a_{0}$ of degree $n$ over the Galois field ${\rm GF}(2)$ and $n$-bit random initial vector. The LFSR-based Toeplitz hashing operation can be written as $h_{\vec{p},\vec{s}}(M)$= $H_{nm} \cdot \vec{M}=\vec{Hash}$,  where  $\vec{p}=(a_{n-1},a_{n-2},...,a_{1},a_{0})$ represents an irreducible polynomial and $\vec{s}=(b_{n},b_{n-1},...,b_2,b_1)^T$ is the initial column vector. $\vec{p}$ and $\vec{s}$ are random and determine the Toeplitz matrix $H_{nm}$ with $n$ rows and $m$ columns. $\vec{M}=(M_{1},M_{2}...M_{m})^{T}$ is the message and known by Eve in the form of an $m$-bit column vector. Note that the tag of the authentication communication is encrypted by one-time pad i.e., $\vec{Tag}=h_{\vec{p},\vec{s}}(\vec{M})\oplus \vec{Key}$. In each authentication communication round, Eve cannot obtain any information of the LFSR-based Toeplitz hash function and the hash value due to $\vec{Key}$ is random.
Therefore the initial vector $\vec{s}$ and irreducible polynomial $\vec{p}$ can be reused to generated the LFSR-based Toeplitz matrix $H_{nm}$ in many authentication communication round.
As a counterpart, in our random information negotiation, $\vec{R}_{n}(\vec{x}_{a})$ is totally random (provable quantum one-way function does not leak any information) and updated in each session.

\section{IV. Secret key rate}

\subsection{A. Proof of zero phase error}

Here, we first introduce entanglement-based virtual protocol 2 in the main text, where Alice and Bob do not directly generate a weak coherent state but rather prepare an entangled state between the qubit and the signal optical mode.

(i) Alice exploits a light source and phase modulator to generate a signal optical pulse $\ket{\sqrt{\mu}e^{\textbf{i}\theta_{a}}}$. She utilizes a Hadamard gate to qubit $A$ to generate the state $\ket{+x}_{A}=\frac{1}{\sqrt{2}}(\ket{+z}_{A}+\ket{-z}_{A})$. Then, Alice exploits a controlled-phase gate $\ket{+z}\bra{+z}\otimes \hat{\textbf{I}}+\ket{-z}\bra{-z}\otimes e^{\textbf{i}\pi \hat{n}}$ to qubit $A$ as a control and signal optical pulse $a$ as a target to generate the entangled state $\ket{\varphi}_{Aa}^{\theta_{a}}=\frac{1}{\sqrt{2}}\left(\ket{+z}_{A}\ket{\sqrt{\mu}e^{\textbf{i}\theta_{a}}}_{a}+\ket{-z}_{A}\ket{-\sqrt{\mu}e^{\textbf{i}\theta_{a}}}_{a}\right)$.
The random global phase $\theta_{a}\in\{0,2\pi/m,4\pi/m,\ldots,2\pi(m-1)/m\}$ is determined by the random bit substring $\vec{x}_{a}\in\{0,1\}^{\log_{2}m}$. There are $m$ global phases and each $\log_{2}m$-bit map to a global phase. Bob does the same. Instead of the fixed mapping rule, the global phase is determined by random mapping rule. The random mapping rule is shared between Alice and Bob through one-time pad encryption by consuming $m\log_{2}m$ bits of the pre-shared secret key.
Alice (Bob) keeps the qubits $A$ ($B$) in quantum memory. Alice and Bob repeat step (i) for $N$ rounds.

(ii) Alice (Bob) utilizes another laser to generate the reference pulse. Alice (Bob) performs the single-photon interference measurement for the signal optical pulse $a$ ($b$) and reference pulse by using a beam splitter and two detectors, $D_{aL}$ and $D_{aR}$ ($D_{bL}$ and $D_{bR}$).
If and only if one of the $D_{aL}$ and $D_{aR}$ ($D_{bL}$ and $D_{bR}$) detector clicks represents a successful detection event. The number of successful detection events for Alice's signal optical pulse and Bob's signal optical pulse are both $n$ for one PKD session. They announce the successful detection event and the corresponding detector.  Alice (Bob) keeps the corresponding qubit $A$ ($B$) of her (his) successful detection event and discards the others.

(iii) Alice and Bob obtain the data string $\vec{D}=\vec{K}\cdot\textbf{H}_{st}$, where $\vec{D}$ has $t$ bits and $t\geq n\log_{2}m$. Let the first $n\log_{2}m$ bits in $\vec{D}$ constitute the data string $\vec{D}_{n}$. For $n$ successful detection events of Alice, let $\vec{R}_{n}(\vec{x}_{a})\in\{0,1\}^{n\log_{2}m}$ be the random bit string corresponding to Alice's global phases. Alice sends ciphertext $\vec{R}_{n}(\vec{x}_{a})\oplus \vec{D}_{n}$ to Bob, who decrypts ciphertext with data string $\vec{D}_{n}$ to obtain quantum random numbers $\vec{R}_{n}(\vec{x}_{a})$ and thus acquires the global phase of Alice $\theta_{a}$ according to the same random mapping rule. According to the phase $\theta_{a}$ of each event, Bob rearranges his qubit system $B$ and detector click order to ensure that $\theta_{b}=\theta_{a}$.
If $\{D_{aL},D_{bR}\}$ or $\{D_{aR},D_{bL}\}$ click, Bob performs a bit flip about his qubit $B$. Then, Alice and Bob exploit the $Z$ basis to measure qubit systems $A$ and $B$ to acquire the raw key, respectively. Finally, Alice and Bob obtain the raw key strings $\vec{Z}_{a}$ and $\vec{Z}_{b}$, respectively.

(iv) Bob acquires an estimation $\vec{\textrm{Z}}_{b}$ of $\vec{Z}_{a}$ in the error correction scheme by revealing at most $\lambda$ bits of information. Then, Alice and Bob perform an error verification to ensure identical keys by publishing the $\log_{2}(2/\varepsilon_{\rm cor})$-bit.

(v) Alice and Bob perform privacy amplification by applying a random universal$_2$ hash function to extract length $\ell$ bits of the secret key.

Obviously, Alice's and Bob's local measurements for the qubit system commute with their measurements for optical modes. The entanglement-based protocol is equivalent to the prepare-and-measure protocol since the local $Z$ basis measurement can be securely moved to step (i). If Alice measures $\ket{+z}_{A}$, she records bit 0 and sends an optical pulse $\ket{\sqrt{\mu}e^{\textbf{i}\theta_{a}}}$. If Alice measures $\ket{-z}_{A}$, she records bit 1 and sends an optical pulse $\ket{-\sqrt{\mu}e^{\textbf{i}\theta_{a}}}$. Actually, from the view of Eve, the nonlocal entangled state will be generated if Alice and Bob have identical and confidential random phase information $\theta_{a}=\theta_{b}$. Alice and Bob can exploit the virtual entangled state to extract the secret key. Alice and Bob can measure two-qubit systems $A$ and $B$ by using the $Z$ and $X$ bases, respectively. The results of the $Z$ basis are used as the raw key. The quantum bit error rate of the $X$ basis quantifies the phase error rate. As follows, we will show that the quantum bit error rate of the $X$ basis is always zero. Thus, we do not need to actually measure the $X$ basis and obtain the zero-phase error conclusion.

The four Bell states can be written as
\begin{equation}
\begin{aligned}\label{eq6}
\ket{\phi^{+}}&=\frac{1}{\sqrt{2}}\left(\ket{+z+z}+\ket{-z-z}\right)=\frac{1}{\sqrt{2}}\left(\ket{+x+x}+\ket{-x-x}\right),\\
\ket{\phi^{-}}&=\frac{1}{\sqrt{2}}\left(\ket{+z+z}-\ket{-z-z}\right)=\frac{1}{\sqrt{2}}\left(\ket{+x-x}+\ket{-x+x}\right),\\
\ket{\psi^{+}}&=\frac{1}{\sqrt{2}}\left(\ket{+z-z}+\ket{-z+z}\right)=\frac{1}{\sqrt{2}}\left(\ket{+x+x}-\ket{-x-x}\right),\\
\ket{\psi^{+}}&=\frac{1}{\sqrt{2}}\left(\ket{+z-z}-\ket{-z+z}\right)=\frac{1}{\sqrt{2}}\left(\ket{-x+x}-\ket{+x-x}\right),\\
\end{aligned}
\end{equation}
where $\ket{\pm z}$ and $\ket{\pm x}$ are the eigenstates of the $Z$ and $X$ bases.
The entangled states generated by Alice and Bob with random phases $\theta_{a}$ and $\theta_{b}$ can be written as
\begin{equation}
\begin{aligned}\label{eq6}
\ket{\varphi}_{Aa}^{\theta_{a}}&=\frac{1}{\sqrt{2}}\left(\ket{+z}_{A}\ket{\sqrt{\mu}e^{i\theta_{a}}}_{a}+\ket{-z}_{A}\ket{-\sqrt{\mu}e^{i\theta_{a}}}_{a}\right),\\
\ket{\varphi}_{Bb}^{\theta_{b}}&=\frac{1}{\sqrt{2}}\left(\ket{+z}_{B}\ket{\sqrt{\mu}e^{i\theta_{b}}}_{b}+\ket{-z}_{B}\ket{-\sqrt{\mu}e^{i\theta_{b}}}_{b}\right).\\
\end{aligned}
\end{equation}
Let us consider the special case with $m\rightarrow\infty$ and $\theta_{b}=\theta_{a}=\theta$. The initial joint quantum state between Alice and Bob can be given by
\begin{equation}
\begin{aligned}\label{eq6}
&\rho=\frac{1}{2\pi}\int_{0}^{2\pi}\ket{\psi}_{Aa}^{\theta}\bra{\psi}\otimes\ket{\psi}_{Bb}^{\theta}\bra{\psi}d\theta
=\frac{1}{2\pi}\int_{0}^{2\pi}\hat{P}\left(\ket{\psi}_{Aa}^{\theta}\ket{\psi}_{Bb}^{\theta}\right)d\theta\\
&=\sum_{k=0}^{\infty}\frac{e^{-2\mu}}{2}\left[\frac{(2\mu)^{2k+1}}{(2k+1)!}\hat{P}\left(\ket{\phi^{-}}_{AB}\ket{+}_{ab}^{\otimes 2k+1}+\ket{\psi^{-}}_{AB}\ket{-}_{ab}^{\otimes 2k+1}\right)+\frac{(2\mu)^{2k}}{(2k)!}\hat{P}\left(\ket{\phi^{+}}_{AB}\ket{+}_{ab}^{\otimes 2k}+\ket{\psi^{+}}_{AB}\ket{-}_{ab}^{\otimes 2k}\right)\right]\\
&=\sum_{k=0}^{\infty}e^{-2\mu}\left[\frac{(2\mu)^{2k+1}}{(2k+1)!}\hat{P}\left(\frac{\ket{\phi^{-}}_{AB}\ket{+}_{ab}^{\otimes 2k+1}+\ket{\psi^{-}}_{AB}\ket{-}_{ab}^{\otimes 2k+1}}{\sqrt{2}}\right)+\frac{(2\mu)^{2k}}{(2k)!}\hat{P}\left(\frac{\ket{\phi^{+}}_{AB}\ket{+}_{ab}^{\otimes 2k}+\ket{\psi^{+}}_{AB}\ket{-}_{ab}^{\otimes 2k}}{\sqrt{2}}\right)\right]\\
&=\sum_{k=0}^{\infty}e^{-2\mu}\frac{(2\mu)^{k}}{k!}\rho_{k},
\end{aligned}
\end{equation}
which is the mixture of $\rho_{k}$ with probability $e^{-2\mu}\frac{(2\mu)^{k}}{k!}$. We have $\hat{P}(\ket{x})=\ket{x}\bra{x}$ and $\ket{\pm}_{ab}=\frac{1}{\sqrt{2}}(\ket{10}_{ab}\pm\ket{01}_{ab})=\frac{\hat{a}^{\dagger}\pm \hat{b}^{\dagger}}{\sqrt{2}}\ket{00}_{ab}$, which is a superposition single-photon state with $a$ and $b$ modes. $\ket{\pm}_{ab}^{\otimes k}=\frac{1}{\sqrt{2^{k}k!}}\left(\hat{a}^{\dag}\pm \hat{b}^{\dag}\right)^{k}\ket{00}_{ab}$ is the $k$-photon state, i.e.,  $k$ identical photons with state $\ket{\pm}_{ab}$. The density matrix of $\rho_{k}$ can be written as
\begin{equation}\label{eqs38}
    \rho_{k}=
   \begin{cases}
   \hat{P}\left(\frac{\ket{\phi^{-}}_{AB}\ket{+}_{ab}^{\otimes k}+\ket{\psi^{-}}_{AB}\ket{-}_{ab}^{\otimes k}}{\sqrt{2}}\right),&\mbox{if $k$ is odd},\\
   \\
  \hat{P}\left(\frac{\ket{\phi^{+}}_{AB}\ket{+}_{ab}^{\otimes k}+\ket{\psi^{+}}_{AB}\ket{-}_{ab}^{\otimes k}}{\sqrt{2}}\right),&\mbox{if $k$ is even}.
   \end{cases}
  \end{equation}
From Eq.~\eqref{eqs38}, no matter what we do with the optical signal modes $a$ and $b$, we can see that the measurement outcomes of the $X$ basis by Alice's qubit $A$ and Bob's qubit $B$ are always perfect anticorrelation (positive correlation) if $k$ is odd (even).
Note that the global phase $\theta_{a}=\theta$ is random due to our first observation and the second observation. Each state $\rho_{k}$ can be regarded as the tagged state, where Alice and Bob can know with certainty what joint quantum state they are sending in each round. After announcing the successful detection of Alice's signal optical pulse and Bob's signal optical pulse, considering that Alice and Bob measure their kept qubit systems on the $X$ basis instead of the $Z$ basis, the quantum bit error rate is always zero for each state $\rho_{k}$. Therefore, the phase error rate must be zero even though Alice and Bob do not perform the $X$ basis measurement.

Now, we can easily generalize the conclusion to a finite phase number $m$ and an arbitrary phase difference $\delta\theta$. For a given $\delta\theta=\theta_{b}-\theta_{a}$, the initial joint quantum state between Alice and Bob can be written as
\begin{equation}
\begin{aligned}\label{}
\rho^{\delta\theta}&=\frac{1}{m}\sum_{\theta_{a}}\ket{\psi}_{Aa}^{\theta_{a}}\bra{\psi}\otimes\ket{\psi}_{Bb}^{\theta_{a}+\delta\theta}\bra{\psi}\\
&=\sum_{k=0}^{m-1}P^{2\mu}_{m}(k)\rho_{\lambda_{k}}^{\delta\theta},
\end{aligned}
\end{equation}
which is the mixture of $\rho_{\lambda_{k}}^{\delta\theta}$ with probability $P^{2\mu}_{m}(k)=e^{-2\mu}\sum_{l=0}^{\infty}\frac{(2\mu)^{lm+k}}{(lm+k)! }$. The density matrix of $\rho_{\lambda_{k}}^{\delta\theta}$ can be given by
\begin{equation}\label{eqs40}
\rho_{\lambda_{k}}^{\delta\theta}=
\begin{cases}
\hat{P}\left(\frac{\ket{\phi^{-}}_{\tilde{a}\tilde{b}}\ket{\lambda_{k}^{+\delta\theta}}_{ab}
+\ket{\psi^{-}}_{\tilde{a}\tilde{b}}\ket{\lambda_{k}^{-\delta\theta}}_{ab}}{\sqrt{2}} \right),  &\mbox{if $k$ is odd},\\
\\
\hat{P}\left(\frac{\ket{\phi^{+}}_{\tilde{a}\tilde{b}}\ket{\lambda_{k}^{+\delta\theta}}_{ab}+\ket{\psi^{+}}_{\tilde{a}\tilde{b}}\ket{\lambda_{k}^{-\delta\theta}}_{ab}}{\sqrt{2}}\right), &\mbox{if $k$ is even},
\end{cases}
\end{equation}
where the quantum state $\ket{\lambda_{k}^{\pm\delta\theta}}_{ab}$ is denoted as
\begin{equation}
\begin{aligned}
\ket{\lambda_{k}^{\pm\delta\theta}}_{ab}&=\frac{e^{-\mu}}{\sqrt{P^{2\mu}_{m}(k)}}\sum_{l=0}^{\infty}\frac{(\sqrt{2\mu})^{lm+k}}{\sqrt{(lm+k)! }}\ket{\pm\delta\theta}_{ab}^{\otimes lm+k}.\\
\end{aligned}
\end{equation}
Quantum state $\ket{\pm\delta\theta}_{ab}^{\otimes k}=\frac{1}{\sqrt{2^{k}k! }}\left(\hat{a}^{\dag}\pm e^{\textbf{i}\delta\theta}\hat{b}^{\dag}\right)^{k}\ket{00}_{ab}$ represents the $k$ identical photon with state $\ket{\pm \delta\theta}_{ab}=\frac{1}{\sqrt{2}}(\ket{10}_{ab}\pm e^{\textbf{i}\delta\theta}\ket{01}_{ab})$. Like in the special case, each state $\rho_{\lambda_{k}}^{\delta\theta}$ can be regarded as the tagged state due to our first and second observations. From Eq.~\eqref{eqs40} and the above argument, the phase error rate is also zero even if Eve carries out any attacks. Finally, since the phase difference $\delta\theta$ can be arbitrary and even known by Eve, we conclude that the phase error rate of the PKD protocol is always zero.

\subsection{B. Universally composable framework}

Here, we note that our PKD protocol meets the universally composable framework~\cite{muller2009composability}. For each PKD session, our protocol either outcomes a pair of key strings $\vec{S}_{a}$ and $\vec{S}_{b}$ with $|\vec{S}_{a}|=|\vec{S}_{b}|=\ell$ or aborts. For a secure protocol, the key strings should satisfy the correctness and secrecy criteria as follows. Specifically, a protocol is $\varepsilon_{\rm tot}$-secure if it is $\varepsilon_{\textrm{cor}}$-correct and $\varepsilon_{\textrm{sec}}$-secret with $\varepsilon_{\textrm{cor}}+\varepsilon_{\textrm{sec}}\leq\varepsilon_{\rm tot}$.

{Correctness criterion}: a protocol is $\varepsilon_{\textrm{cor}}$-correct if the error-verification step is passed, where we have ${\rm Pr}[\vec{S}_{a}\neq \vec{S}_{b}]\leq\varepsilon_{\rm cor}$, i.e., the probability that two key strings are not identical does not surpass $\varepsilon_{\textrm{cor}}$.

{Secrecy criterion}: an ideal protocol in which the generated key string $\vec{S}_{a}$ of Alice is independent of Eve's system $\rho_{E}$, i.e., the joint classical-quantum state between Alice and Eve
\begin{equation}
\begin{aligned}
\rho_{AE}^{\rm ideal}&=U_{A}\otimes\rho_{E}\\
&=\frac{1}{2^{\ell}}\sum_{\vec{s}_{a}}\ket{\vec{s}_{a}}\bra{\vec{s}_{a}}\otimes\rho_{E},
\end{aligned}
\end{equation}
where $\vec{s}_{a}\in\{0,1\}^{\ell}$ is the bit value and $U_{A}=\frac{1}{2^{\ell}}\sum_{\vec{s}_{a}}\ket{\vec{s}_{a}}\bra{\vec{s}_{a}}$ is the uniform mixture of all possible values of the key string $\vec{S}_{a}$. The ideal case is never perfectly satisfied and should be the nonideal case in the experiment. The corresponding state can be defined as $\rho_{AE}=\sum_{\vec{s}_{a}}p_{\vec{s}_{a}}\ket{\vec{s}_{a}}\bra{\vec{s}_{a}}\otimes \rho_{E}^{\vec{s}_{a}}$. A protocol is $\varepsilon_{\rm sec}$-secret after the privacy amplification step if we have
\begin{equation}
\begin{aligned}\label{eqll}
\frac{1}{2}(1-p_{\rm about})\|\rho_{AE}-U_{A}\otimes\rho_{E}\|_{1} \leq \varepsilon_{\rm sec},
\end{aligned}
\end{equation}
where $p_{\rm abort}$ is the probability that the protocol will abort.

\subsection{C. Entropic uncertainty relation}

Note that the prepare-and-measure PKD is equivalent to the entanglement-based protocol. The entropic uncertainty relations~\cite{tomamichel2012tight} can be utilized to estimate the smooth min-entropy of the raw key conditioned on Eve's information.
Up to the error-correction step is complete, let $\textbf{E}'$ denote all the information that is acquired from Eve about the raw key $\vec{Z}_{a}$ of Alice.
Let $H_{\rm min}^{\epsilon}(\vec{Z}_{a}|\textbf{E}')$ be the smooth min-entropy that quantifies the average probability of Eve correctly guessing $\vec{Z}_{a}$.
According to the quantum leftover hash lemma, one can exploit a random universal$_{2}$ hash function to $\vec{Z}_{a}$, and a $\varepsilon_{\rm sec}$-secret key of length $\ell$ from $\vec{Z}_{a}$ can be extracted
\begin{equation}
\begin{aligned}
2\epsilon+\frac{1}{2}\sqrt{2^{\ell-H_{\rm min}^{\epsilon}(\vec{Z}_{a}|\textbf{E}')}}\leq\varepsilon_{\rm sec},
\end{aligned}
\end{equation}
Let $\varepsilon_{\rm PA}=\frac{1}{2}\sqrt{2^{\ell-H_{\rm min}^{\epsilon}(\vec{Z}_{a}|\textbf{E}')}}$ be a security parameter related to privacy amplification; the secret key of length $\ell$ is
\begin{equation}
\begin{aligned}\label{eqs45}
\ell=\bigg\lfloor H_{\rm min}^{\epsilon}(\vec{Z}_{a}|\textbf{E}')-2\log_{2}\frac{1}{2\varepsilon_{\rm PA}}\bigg\rfloor.
\end{aligned}
\end{equation}

During the error-correction step and error-verification step, a total of $\lambda+\log_{2}(2/\varepsilon_{\rm cor})$-bit will be published to Eve.
Using the chain-rule inequality for smooth entropies, the smooth min-entropy can be bounded by
\begin{equation}
\begin{aligned}\label{eqs46}
H_{\rm min}^{\epsilon}(\vec{Z}_{a}|\textbf{E}')\geq H_{\rm min}^{\epsilon}(\vec{Z}_{a}|\textbf{E})-\lambda-\log_{2}\frac{2}{\varepsilon_{\rm cor}},
\end{aligned}
\end{equation}
where $\textbf{E}$ is the information learned by Eve before error correction. We have proven that the phase error rate $\phi_{z}$ is strictly zero. According to the uncertainty relation for smooth entropies, the smooth min-entropy $H_{\rm min}^{\epsilon}(\vec{Z}_{a}|\textbf{E})$ can be written as
\begin{equation}
\begin{aligned}\label{eqs47}
H_{\rm min}^{\epsilon}(\vec{Z}_{a}|\textbf{E})&\geq n[1-h(\phi_{\rm z})]=n.
\end{aligned}
\end{equation}
Finally, we let $\epsilon=\varepsilon_{\rm PA}=\varepsilon_{\rm sec}/3$ and $2\epsilon+\varepsilon_{\rm PA}=\varepsilon_{\rm sec}$, and we can obtain the secret key rate by combining Eqs.~\eqref{eqs45}-\eqref{eqs47},
\begin{equation}
\begin{aligned}
\ell=n-\lambda-\log_{2}\frac{2}{\varepsilon_{\rm cor}}-2\log_{2}\frac{3}{2\varepsilon_{\rm sec}},
\end{aligned}
\end{equation}
where $n$ is the length of the raw key and $\lambda=nfh(E)$ is the revealed information in the error-correction step.

\section{V. Simulation formula}

When Alice prepares the optical pulse pairs $\ket{\sqrt{\mu}e^{\textbf{i}(\theta_{a}+r_{a}\pi)}}_{a}\otimes\ket{\sqrt{\mu}}_{a}$ and performs the single-photon measurement, the gains $Q_{aL}$ and $Q_{aR}$ can be given by
\begin{equation}
\begin{aligned}\label{}
Q_{aL}=\left\{1-(1-p_{d})e^{-\mu\eta_{d}[1+\cos(\theta_{a}+r_{a}\pi)]}\right\}(1-p_{d})e^{-\mu\eta_{d}[1-\cos(\theta_{a}+r_{a}\pi)]},\\
\end{aligned}
\end{equation}
and
\begin{equation}
\begin{aligned}\label{}
Q_{aR}=\left\{1-(1-p_{d})e^{-\mu\eta_{d}[1-\cos(\theta_{a}+r_{a}\pi)]}\right\}(1-p_{d})e^{-\mu\eta_{d}[1+\cos(\theta_{a}+r_{a}\pi)]},\\
\end{aligned}
\end{equation}
where $Q_{aL}$ ($Q_{aR}$) represents that detector $D_{aL}$ ($D_{aR}$) has a click and $D_{aR}$ ($D_{aL}$) does not have a click. $p_{d}$ and $\eta_{d}$ are the dark count rate and detection efficiency of the detector, respectively.

When Bob prepares the optical pulse pairs $\ket{\sqrt{\mu}e^{\textbf{i}(\theta_{b}+r_{b}\pi)}}_{b}\otimes\ket{\sqrt{\mu}}_{b}$ and performs the single-photon measurement, the gains $Q_{bL}$ and $Q_{bR}$ can be given by
\begin{equation}
\begin{aligned}\label{}
Q_{bL}=\left\{1-(1-p_{d})e^{-\mu\eta_{d}[1+\cos(\theta_{b}+r_{b}\pi)]}\right\}(1-p_{d})e^{-\mu\eta_{d}[1-\cos(\theta_{b}+r_{b}\pi)]},\\
\end{aligned}
\end{equation}
and
\begin{equation}
\begin{aligned}\label{}
Q_{bR}=\left\{1-(1-p_{d})e^{-\mu\eta_{d}[1-\cos(\theta_{b}+r_{b}\pi)]}\right\}(1-p_{d})e^{-\mu\eta_{d}[1+\cos(\theta_{b}+r_{b}\pi)]},\\
\end{aligned}
\end{equation}
where $Q_{bL}$ ($Q_{bR}$) represents that detector $D_{bL}$ ($D_{bR}$) has a click and $D_{bR}$ ($D_{bL}$) does not have a click.

For one PKD session, the total number of successful detection events can be given by
\begin{equation}
\begin{aligned}\label{}
n&=\frac{N}{m}\sum_{\theta_{a}=0}^{2\pi(m-1)/m}(Q_{aL}+Q_{aR})=\frac{N}{m}\sum_{\theta_{b}=0}^{2\pi(m-1)/m}(Q_{bL}+Q_{bR})
\approx\frac{N}{2\pi}\int_{0}^{2\pi}(Q_{aL}+Q_{aR})d\theta_{a},\\
&=2N\left[(1-p_{d})e^{-\mu\eta_{d}}I_{0}(\mu\eta_{d})-(1-p_{d})^{2}e^{-2\mu\eta_{d}}\right],
\end{aligned}
\end{equation}
where $I_{0}(x)$ is the modified Bessel function of the first kind. After the raw key rearrangement, the global phases of the Alice signal pulse and the Bob signal pulse are always the same as $\theta_{a}=\theta_{b}=\theta$. Therefore, we have $Q_{L}=\left[1-(1-p_{d})e^{-\mu\eta_{d}(1+\cos\theta)}\right](1-p_{d})e^{-\mu\eta_{d}(1-\cos\theta)}$ and $Q_{R}=\left[1-(1-p_{d})e^{-\mu\eta_{d}(1-\cos\theta)}\right](1-p_{d})e^{-\mu\eta_{d}(1+\cos\theta)}$. An error will occur if $r_{a}=r_{b}$ and detectors $\{D_{aL}, D_{bR}\}$ ($\{D_{aR}, D_{bL}\}$) click or if $r_{a}\neq r_{b}$ and detectors $\{D_{aL}, D_{bL}\}$ ($\{D_{aR}, D_{bR}\}$) click. The results for the discrete phase-randomized cases are almost the same as those for the continuous phase-randomized cases; thus, the bit error rate can be written as
\begin{equation}
\begin{aligned}\label{}
E&=\frac{1}{2\pi}\int_{0}^{2\pi}\frac{2Q_{L}Q_{R}}{(Q_{L}+Q_{R})^{2}}d\theta\\
&=\frac{1}{2\pi}\int_{0}^{2\pi}\frac{2(1-p_{d})^{2}e^{-2\mu\eta_{d}}\left[1-(1-p_{d})e^{-\mu\eta_{d}(1+\cos\theta)}\right]\left[1-(1-p_{d})e^{-\mu\eta_{d}(1-\cos\theta)}\right]}{\left[(1-p_{d})e^{-\mu\eta_{d}}(e^{-\mu\eta_{d}\cos\theta}+e^{\mu\eta_{d}\cos\theta})-2(1-p_{d})^{2}e^{-2\mu\eta_{d}}\right]^{2}}d\theta.
\end{aligned}
\end{equation}
For $\mu=0.1$, $\eta_{d}=0.8$ and $p_{d}=10^{-8}$, the bit error rate is $E\approx25\%$.

